\documentclass[aps,showpacs,preprintnumbers,amsmath,amssymb]{revtex4}
\usepackage{dcolumn}
\usepackage[dvips]{epsfig}

\usepackage{float}
\usepackage{graphics,epsfig}
\usepackage{graphicx}
\usepackage{dcolumn}
\usepackage{bm}

\begin{document}
\baselineskip=0.8 cm
\title{\bf   Periodic orbits around Kerr Sen black holes }

\author{
Changqing Liu$^{1}$\footnote{Electronic address:
lcqliu2562@163.com}, Chikun Ding$^{1}$\footnote{Electronic
address: Chikun\_Ding@huhst.edu.cn}, Jiliang
Jing$^{2}$\footnote{Electronic address: jljing@hunnu.edu.cn}}

\affiliation{1) Department of Physics, Hunan University of
Humanities Science and Technology, Loudi, Hunan 417000, P. R. China}

\affiliation{2) Department of Physics, and Key Laboratory of Low
Dimensional Quantum Structures and Quantum Control of Ministry of
Education, Hunan Normal University,  Changsha, Hunan 410081, P. R.
China}
\begin{abstract}
\baselineskip=0.6 cm
\begin{center}
{\bf Abstract}
\end{center}
 We investigate periodic orbits and zoom-whirl behaviors around a Kerr Sen black hole with a rational number
 $q$ in terms of three integers $(z,w,v)$, from
which one can immediately read off the number of leaves
(or zooms), the ordering of the leaves, and the number of
whirls. The characteristic of
 zoom-whirl periodic orbits is the precession of multi-leaf orbits in the
strong-field regime. This feature is analogous to the counterpart in the Kerr space-time. Finally, we analyze the impact of the charge parameter $b$ on the
zoom-whirl periodic orbits.
Compared to the periodic orbits around the Kerr black hole, it is found that
 typically lower energies are required for the same orbits in the Kerr Sen  black hole.
\end{abstract}

 \maketitle
\newpage
\section{Introduction}

Periodic orbits have played a crucial role in the treatment of
some difficult problems in celestial mechanics, including the motions of planetary satellites,
the long term stability of the solar system, and motion in galactic potential.
It is fact that the relativistic
precession of Mercury¡¯s perihelion in the weak
field is around a star. In the strong-field, perihelion precession
in the equatorial plane of a black hole can result in
zoom-whirl orbits for which the precession is so great at
closest approach that the particle executes multiple circles
before falling out to apastron again. The Laser Interferometer Gravitational-wave Observatory (LIGO) \cite{gw1,gw2,gw3} and VIRGO collaborations reported the observation of gravitational-wave
signal corresponding to the inspiral and merger of two black holes is also relevant to this relativistic trajectories. In a series of papers \cite{Levin,Levin1,Levin2,Levin3,Levin4}, Levin et al, proposed a classification of the zoom-whirl structure of the periodic orbits around black hole by using Kerr
geodesics \cite{sha2,Chandra,wilkins,Hughes2,Barausse,Ryuichi,GK,Celestial,Mino:2003yg,Drasco:2004,Kostas_Review} with a rational number
 $q$ in terms of three integers $(z,w,v)$
\begin{align}
q=w+\frac{v}{z}
\end{align}
where $w$ counts the number of whirls, $z$ counts the number
of leaves, and $v$ indicates the order in which the leaves are traced out.
The rational number $q$ explicitly measures the degree of perihelion precession beyond the ellipse as well as the topology of the
orbit. This classification is applied to black hole pairs, they found that zoom-whirl behavior is ubiquitous in comparable
mass binary dynamics and entirely quantifiable through
the spectrum of rational. This  zoom-whirl behavior is also found in the Reissner-Nordstr$\ddot{o}$m black hole \cite{Levin3} and spherically symmetric naked
singularity \cite{gzBabar}, Kehagias-Sfetsos black hole \cite{shaowen}. Furthermore, periodic orbits are generalized from the equatorial taxonomy
to fully generic 3D Kerr motion \cite{Levin4}.

The Kerr-Sen black hole (KSBH) solution \cite{sen} is a charged and rotating solution
 in the low energy limit of heterotic string theory and is also characterized
by mass, electric charges, and angular momentum, which are similar to those of the Kerr Newman
black hole. Some distinguishable properties and various aspects of particle motion
\cite{Blaga:2001wt,Pradhan:2015yea, Hioki2008zw,Koga1995bs, Ghez2012qn,Furu2004,Siahaan2015xna, Dastan:2016bfy} in those space-times have been studied.
Based on a topological taxonomy of periodic orbit, in this paper, we
will use Levin's \cite{Levin} classification scheme to investigate the zoom-whirl behavior and
orbital dynamics in the equatorial plane of the KSBH.  We will use specific features of the periodic orbits to distinguish KSBH from
Kerr black hole.

The paper is organized as follows: In Sec. II, we first derive the relevant geodesic equations of KSBH using the Hamiltonian formulation. In Sec. III, we investigate the innermost bound and stable circular orbits, as well as a qualitative analysis of the effective potential. In Sec.IV,  the energy of zoom-whirl periodic orbits in the KSBH is studied.
Finally, we end the paper
with a summary.

\section{The time-like geodesic equations in the  Kerr Sen black hole }

In Ref. \cite{sen}, Sen obtained a four dimensional solution that describes a rotating and electrically
charged massive body in the low energy heterotic string field theory. In the
Boyer-Lindquist coordinates, the Kerr-Sen metric can be
rewritten as
\begin{eqnarray}
ds^{2}=&-&\left(1-\frac{2Mr}{\rho^{2}}\right)dt^{2}+\rho^{2}\left(\frac{dr^{2}}{\Delta}+d\theta^{2}\right)-\frac{4Mra\sin^{2}{\theta}}{\rho^{2}}dt{d\phi} \nonumber  \\  &+&\left(r(r+b)+a^{2}+\frac{2Mra^{2}\sin^{2}{\theta}}{\rho^{2}} \right)\sin^{2}{\theta}d\phi^{2} \label{LineElement},
\end{eqnarray}
where the functions $\Delta$ and $\rho^{2}$ are given by
\begin{eqnarray}
    \Delta &=& r(r+b)-2Mr+a^{2}, \\
    \rho^{2} &=& r(r+b)+a^{2}\cos^{2}{\theta}.
\end{eqnarray}
Here $M$ is the mass of the black hole, $a$ is the specific angular momentum of the black hole, $b=Q^{2}/M$, $Q$ being the electrical charge of the black hole. In the particular case $b=0$, the above solution is reduced to the Kerr one. The event horizon of the KSBH is located at $r_{H}=\frac{2M-b+\sqrt{(2M-b)^{2}-4a^{2}}}{2}$.

The Hamiltonian of a time-like particle propagating along geodesics in a Kerr Sen black hole can be expressed as
\begin{equation}
H(x_i, p_i) =\frac{1}{2}g^{\mu\nu}(x)p_{\mu}p_{\nu}=
\frac{\Delta}{2\rho^2}p_r^2+\frac{1}{2\rho^2}p_\theta^2+f(r,\theta,p_t,p_\varphi)=-\frac{\mu^2}{2},
\end{equation}
where $\mu$ is the mass of particle. It is easy to obtain two conserved quantities: the energy $E$ and  angular momentum $L$  of the test particles with the following forms
\begin{eqnarray}
\label{EL}
E=-p_{t}=-g_{tt}\dot{t}-g_{t\varphi}\dot{\varphi},\;\;\;\;\;\;\;\;\;\;\;\;\;\;
L=p_{\varphi}=g_{\varphi\varphi}\dot{\varphi}+g_{\varphi t}\dot{t},
\end{eqnarray}

The first integral from the geodesic equations in case of the KSBH  are calculated as follows \citep{Blaga:2001wt,Pradhan:2015yea, Hioki2008zw,Koga1995bs, Ghez2012qn,Furu2004,Siahaan2015xna, Dastan:2016bfy}. Following the procedure in Ref \cite{Levin}, we will convert the first integral equations into
 Hamiltonian formulation to avoid the numerical difficulties and smoothly plot the time-like
zoom-whirl orbits.
With the help of Hamilton's
equations
\begin{equation}
\label{D-Eq} \frac{dx^i}{d \lambda} = \frac{\partial H}{\partial
p_i}  \, , \; \; \frac{dp_i}{d \lambda} = - \frac{\partial
H}{\partial x^i} \, ,
\end{equation}
 the equations of the time-like particle motion become as,
\begin{eqnarray}
\label{eom}
\dot{r} & =& \frac{\Delta}{\rho^2}p_{r}\label{geeoss1} , \\
 \dot{p}_{r} & = &
-\left (\frac{\Delta}{2\rho^2}\right )'p_{r}^{2} -
\left (\frac{1}{2\rho^2}\right )'p_{\theta}^{2} + \left (\frac{R +
 \Delta\Theta}{2\Delta\rho^2}\right )' \label{geeoss2}, \\
\dot{\theta}& = & \frac{1}{\rho^2}p_{\theta}
 ,\\
 \dot{p}_{\theta} & = &
-\left (\frac{\Delta}{2\rho^2}\right )^{\theta}p_{r}^{2} -
 \left (\frac{1}{2\rho^2}\right )^{\theta}p_{\theta}^{2} + \left (\frac{R +
 \Delta\Theta}{2\Delta\rho^2}\right )^{\theta}  \label{geeoss3},\\
  \dot{t} & = & \frac{1}{2\Delta\rho^2} \frac{\partial (R+\Delta\Theta)}{\partial
E}\label{geeoss4}, \\
\dot{\varphi} & = & -\frac{1}{2\Delta\rho^2}
  \frac{\partial(R +\Delta\Theta)}{\partial
L}, \label{geeoss5}
\end{eqnarray}
with
\begin{eqnarray}
\label{jfc}
R(r)&=&-\Delta[r(r+b)+Q+(aE-L)^{2}]+[aL-(r(r+b)+a^{2})E]^{2},
\\
\label{thfc}
\Theta(\theta)&=&Q-\cos^{2}\theta
\bigg(\frac{L^{2}}{\sin^{2}\theta}+a^{2}(1-E^{2})\bigg),
\end{eqnarray}
where the superscripts $'$ and $\theta$ denote differentiation with
respect to $r$ and $\theta$, respectively. The quantity $Q$ is the generalized Carter constant related to the constant of separation $K$ by $Q=K-(aE-L)^2$.
In this paper we  only deal with the motion of bounded time-like $(\mu=1)$ particles in the equatorial plane, for which motion lies in the
4D hypersurface defined by $\theta=\pi/2$, and on which $Q=0$.

\section{Bound on angular momentum $L$   }
 As mentioned in Ref \cite{Levin}, in order to have a sufficiently rich variety of zoom--whirl periodic orbits, the angular momentum $L$ of the particle should satisfies
 \begin{eqnarray}
L_{ISCO}<L<L_{IBCO},
\end{eqnarray}
where ISCO stands for ``Innermost Stable Circular Orbit'' and IBCO for ``Innermost Bound Circular Orbit''.
$L_{ISCO}$ is the lowest value of $L$ for which the potential has
a local minimum. For $L < L _{ISCO}$ , all orbits
will plunge into the black hole, so $L_{ISCO}$ sets the lower
limit on bound orbits. $L_{IBCO}$ marks the first appearance
of an unstable circular orbit that is energetically bound.
It sets the upper limit only in the sense that we expect
to see the most zoom-whirl behavior. From the geodesics, the conditions to determine
\begin{figure}[ht]
\begin{center}
\includegraphics[scale=0.5]{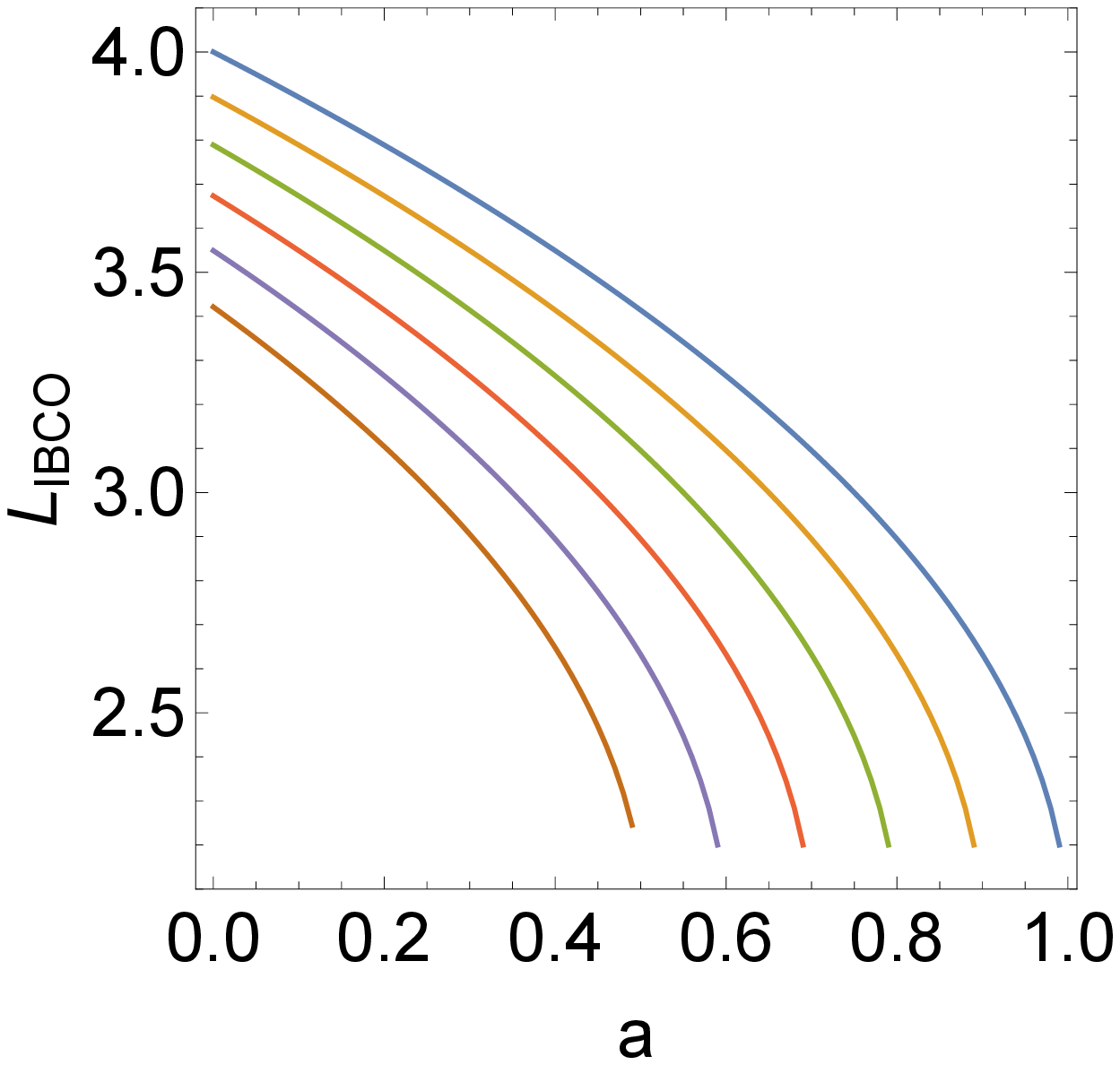},\includegraphics[scale=0.5]{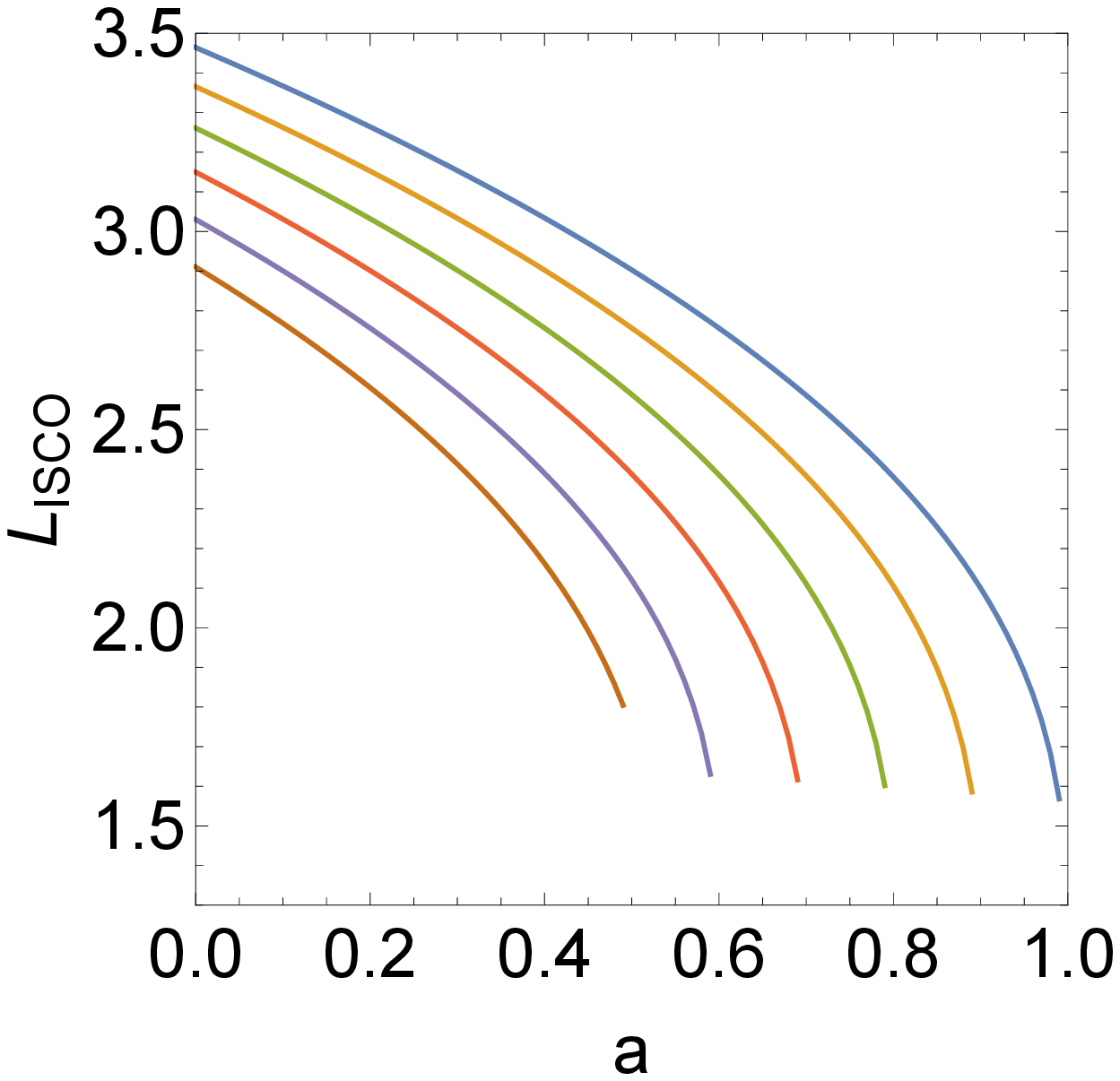}
\caption{Angular momentum $L_{ISCO}$ and  $L_{IBCO}$ vs. $a$ in the  Kerr Sen black hole, with Parameter $b=1, 0.8, 0.6, 0.4, 0.2, 0$ from left to right, here we set $M=1$. \label{figure1}}
\end{center}
\end{figure}
the ISCO are
\begin{eqnarray}
\dot{r}=0,\;\;\;\;\; \ddot{r}=0, \text{and} \;\;\;\;  \dddot{r}=0
\end{eqnarray}
which yield
\begin{eqnarray}
\label{r}
R(r)&=&-\Delta[r(r+b)+(aE-L)^{2}]+[aL-(r(r+b)+a^{2})E]^{2}=0,
\\\nonumber
\label{r1}
R'(r)&=&-2E(2r+b)(aL-E(a^{2}+r(r+b)))-(2r+b)\Delta-((aE-L)^{2}+ r(r+b))\Delta'=0,\\
R''(r)&=&-4E(aL-E(a^{2}+r(r+b)))+2E^2(2r+b)^2-2\Delta-(2r+b)\Delta'-((aE-L)^{2}+ r(r+b))\Delta''=0\nonumber.
\end{eqnarray}
For the non-rotating black hole, these equations can be solved simultaneously for $E$ and $L$ to give
\begin{eqnarray}
 L&=&\pm\frac{\sqrt{2} M r (b+r)}{\sqrt{M (b+r) \left(b^2+b (3 r-2 M)+2 r (r-3 M)\right)}},\\
 E&=&\frac{\sqrt{b+2 r} (b-2 M+r)}{\sqrt{(b+r) ((b+r) (b+2 r)-2 M (b+3 r))}}.
\end{eqnarray}
 And the radius of the ISCO is given by
 \begin{eqnarray}
 r_{ISCO}=2M-b+2^{2/3} \sqrt[3]{M^2 (2 M-b)}+\sqrt[3]{2} \sqrt[3]{M (b-2 M)^2}.
\end{eqnarray}
 When $b=0$, one will get $ r_{ISCO}$ for the Schwarzschild black hole
\begin{eqnarray}
 r_{ISCO}/M=6,\quad
 L_{ISCO}/M\mu=2\sqrt{3},\quad
 E_{ISCO}=\frac{2\sqrt{2}}{3}.
\end{eqnarray}
The radius of the IBCO is given from the condition $E=1$ \cite{Bardeen}
 \begin{eqnarray}
 r_{IBCO}=2M-b+\sqrt{2} \sqrt{2 M^2-b M}.
\end{eqnarray}
While when $a\neq0$, no analytical result is available. Nevertheless, we can obtain a numerical solution. The results are listed in Fig. \ref{figure1} for prograde orbit. For the prograde ISCO and IBCO , both the angular momentum $L_{ISCO}$ and  $L_{IBCO}$ decreases with the black hole spin $a$ and the charge parameter $b$.

For a non-spinning  black hole ($a=0$), we can rewrite the radial
equation  as the expression of effective
potential
\begin{equation}
\label{_Veff_eqn}
\frac{1}{2}(\dot r)^2+V_{\rm eff}=\frac{E^2}{2}
\quad,
\end{equation}
with
\begin{equation}
\label{Veff_eqn1_}
V_{\rm eff}=\frac{1}{2}\frac{\Delta~\left(L^2+r(r+b)\right)}{r^2(r+b)^2}
\quad,
\end{equation}
this effective
potential $V_{\rm eff}$ is a different function of $r$ for each
fixed $L$ and is independent of $E$.  The
result is a simple visual way to describe the different types of
allowed motion as $L$ is varied.
However, the effective
potential $V_{\rm eff}$ of the spinning Kerr Sen black hole is dependent of $E$.
We therefore lose the ability to visualize easily the variation of orbits with
energy. A useful pseudo-effective potential \cite{Levin3} is constructed through
the condition $R(r)=0$ as
\begin{equation}
\left. V_{\rm eff}\right |_{\dot r=0}=\frac{E^2}{2} \quad,
\end{equation}
with
\begin{widetext}
\begin{align}
    \label{eq:V_eff}
E=
\frac{\sqrt{r^2 (b+r) \left(a^2+r (b-2 M+r)\right) \left(a^2 (b+2 M+r)+(b+r)
   \left(r (b+r)+L^2\right)\right)}+2 a L M r}{r \left(a^2 (b+2 M+r)+r
   (b+r)^2\right)}.
\end{align}
\quad .
\end{widetext}
\begin{figure}[ht]
\begin{center}
\includegraphics[scale=0.5]{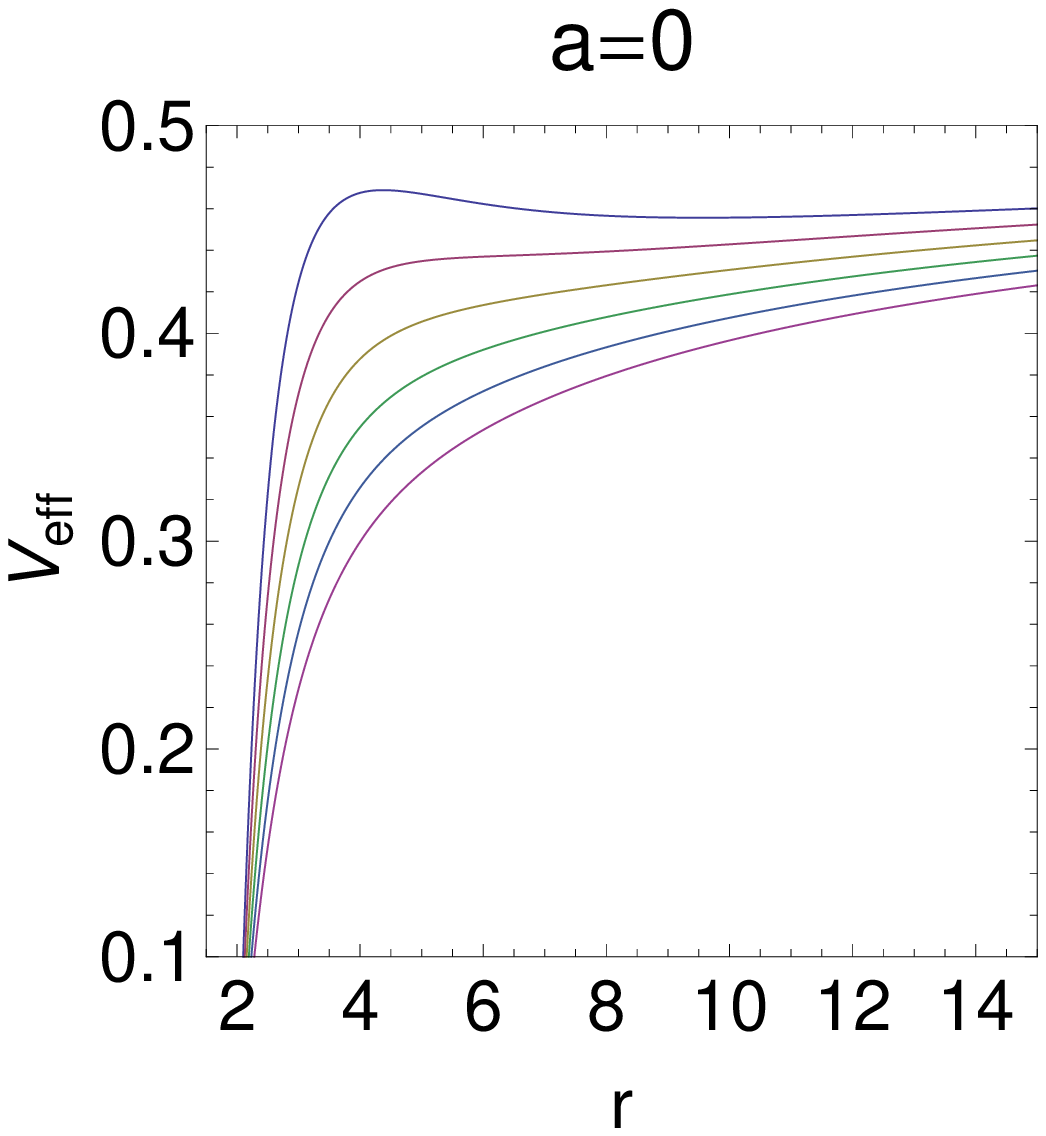},\includegraphics[scale=0.51]{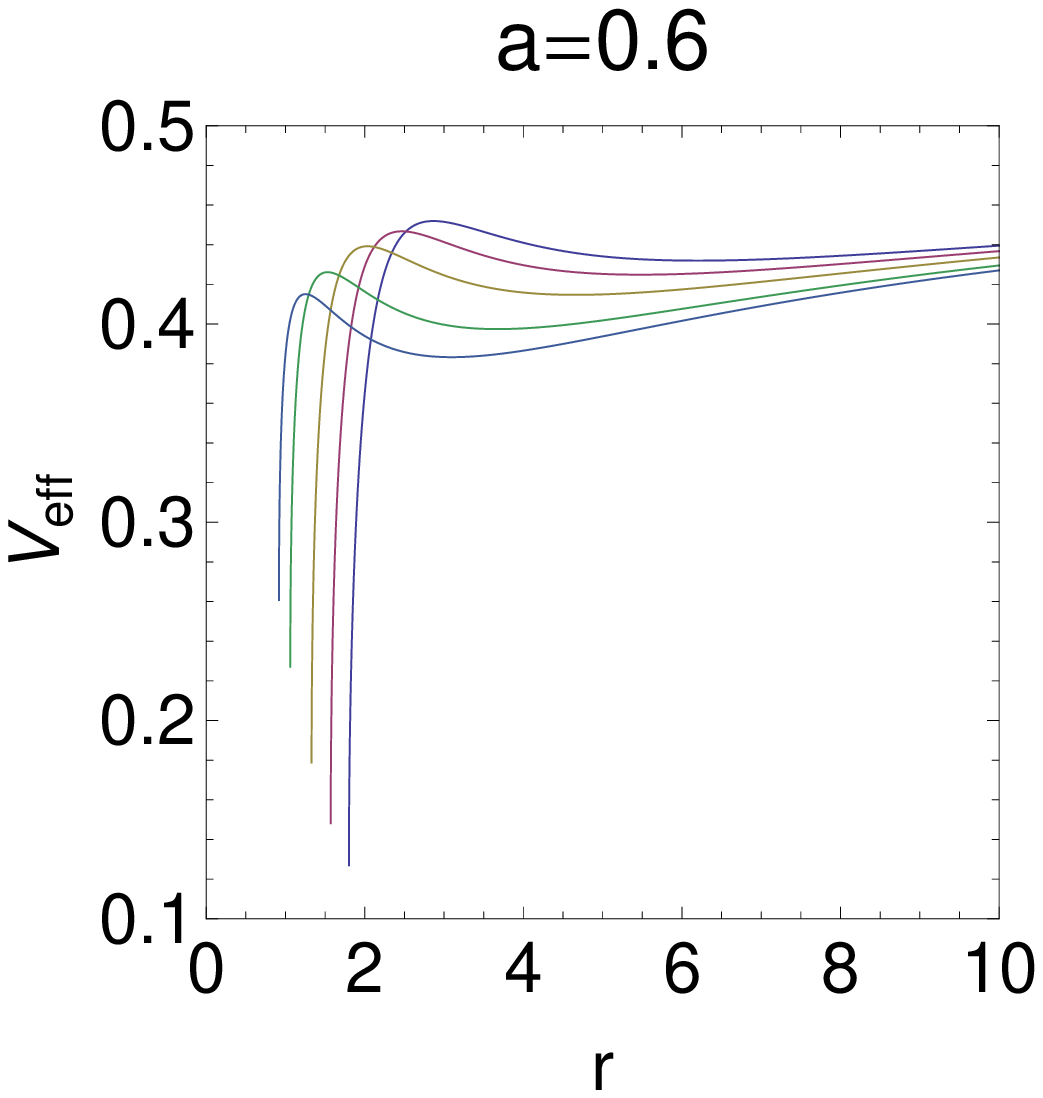}
\caption{Effective potentials with different the charge parameter $b$ for the corresponding angular momentum $L=L_{av}=\frac{L_{ISCO}+L_{IBCO}}{2}$ in the  Kerr Sen black hole: in figure (A), parameter takes the value $b=0, 0.2, 0.4, 0.6, 0.8, 1$ from top  to bottom; in figure (B),
parameter takes the value $b=0.69, 0.6, 0.4, 0.2, 0$ from left  to right. Here we set $M=1$. \label{vef}}
\end{center}
\end{figure}
 Even if the difference between $E$ and the value of
$V_{\rm eff}$ no longer gives the value of $\dot{r}^2$, this pseudo-effective potential illustrates the change of periodic orbits with
energy.

Figure~\ref{vef} depicts the influence of the charge parameter $b$ to the effective potential. The maximum value of
the effective potential decreases with the increasing of the charge parameter $b$. Notice that
the corresponding angular momentum $L$ of the effective potential takes the value $L_{av}$ \cite{gzBabar} (the average value of $L_{ISCO}$ and $L_{IBCO}$)
\begin{equation}
L_{av}=\frac{L_{ISCO}+L_{IBCO}}{2},
\end{equation}
this would give an appropriate potential well for any parameter $b$  that captures most of the physics
of the bound orbits. In the last section \ref{ener},
the angular momentum $L$ also takes the value $L_{av}$ as we analyze the impact the charge parameter $b$ on the energy of the periodic orbit, in order
to have a sufficiently deep potential well that supports a wider variety of orbits .

\section{Periodic orbits in Kerr Sen black hole }
In this section, we shall study zoom-whirl periodic orbits around the KSBH. We use the taxonomy
of orbit of Levin et al. \cite{Levin,Levin1,Levin2,Levin3} to derive the association between periodic orbits and rational numbers $q$
from the dynamical systems perspective. Any bound orbit may be characterized by two fundamental frequencies--the libration in the radial coordinate,  $\omega_r$ , and the rotation in the angular coordinate,
 $\omega_\varphi$. Zoom-whirl periodic orbit corresponds to trajectories where the ratio of these two frequencies is a rational number $q$ in terms of three integers $(z,w,v)$,
\begin{equation}
q=w+\frac{v}{z}\equiv \frac{\omega_\varphi}{\omega_r}-1 =\frac{\Delta
  \varphi}{2\pi}-1,
\label{qequiv}
\end{equation}
where $\Delta \varphi=\int^{T_r}(d\varphi/dt)dt $ is the equatorial
angle accumulated in one
radial cycle from apastron to apastron. By this definition, we see
that $q$ is the amount an
orbit precesses beyond the closed ellipse. These three quantities $z,w,v$ have a geometric interpretation in terms of the structure of the trajectory, where $z$ is the `zoom' number, $w$ is the number of `whirls', and $v$ is the number of vertices formed by joining the successive apastra of the orbits \cite{Levin}.
Thus the trajectory will close and the particle returns to its initial state within a finite (affine) time, thus executing its prior trajectory repeatedly.
 \begin{figure}[ht]
\begin{center}
\includegraphics[width=4cm]{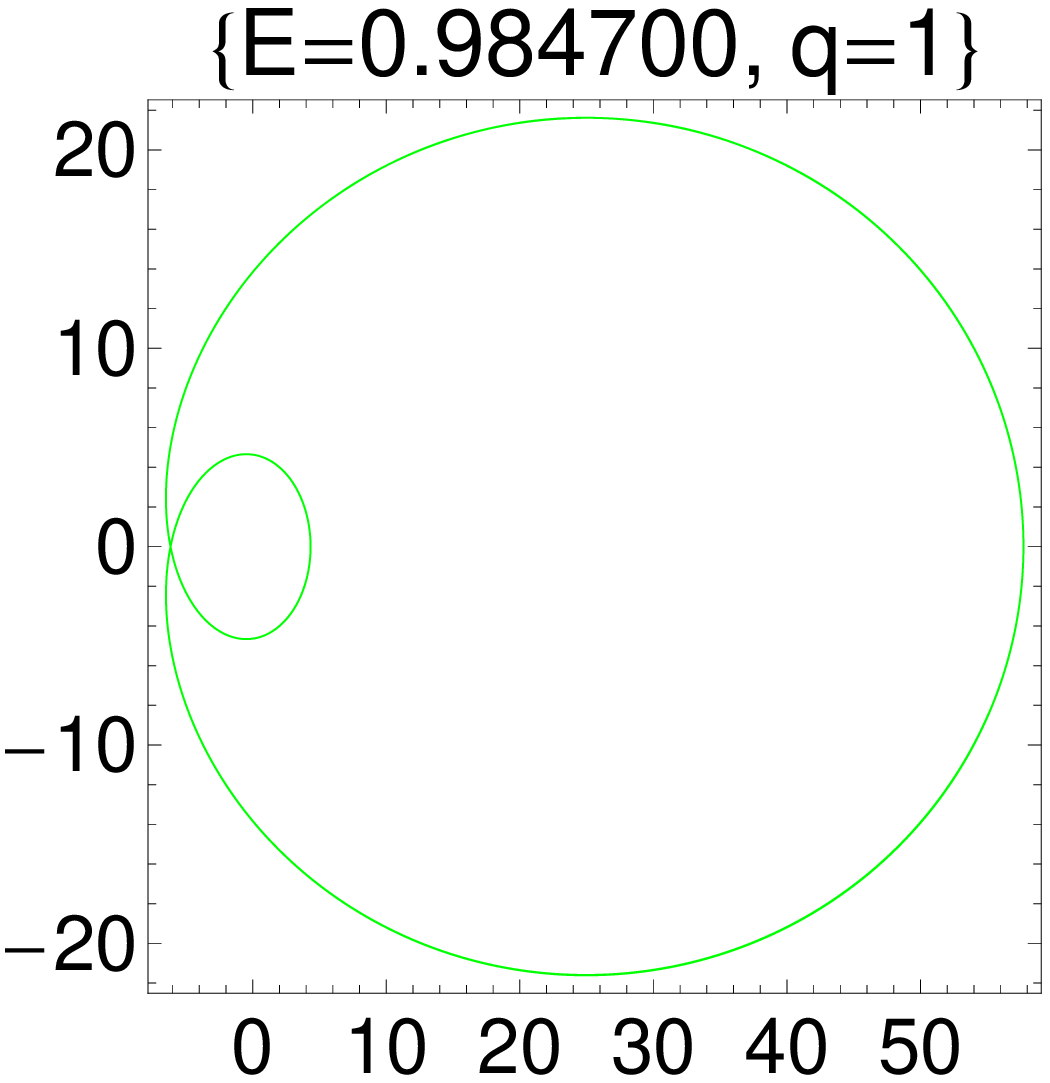}\;\includegraphics[width=4cm]{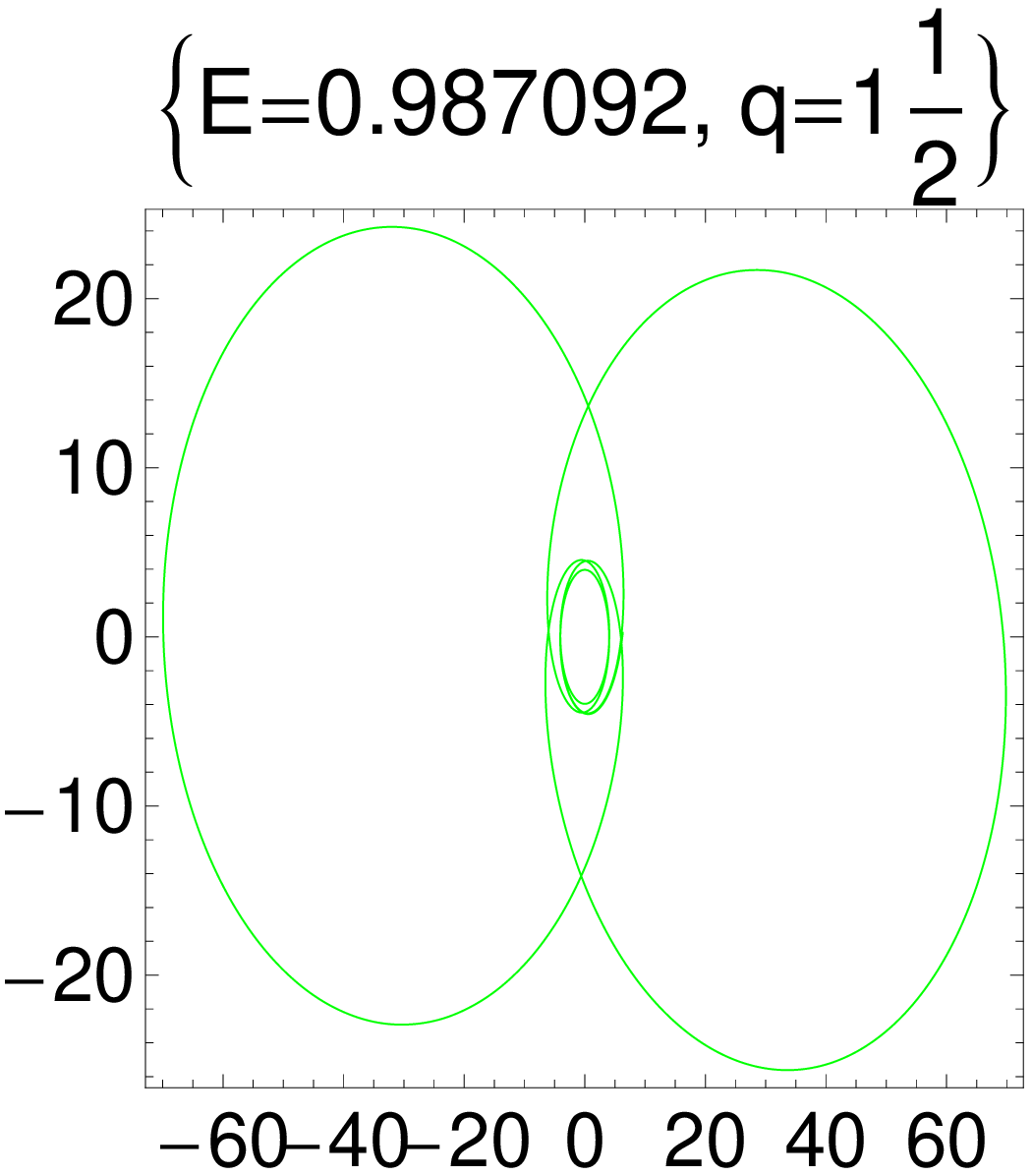}\;\includegraphics[width=4cm]{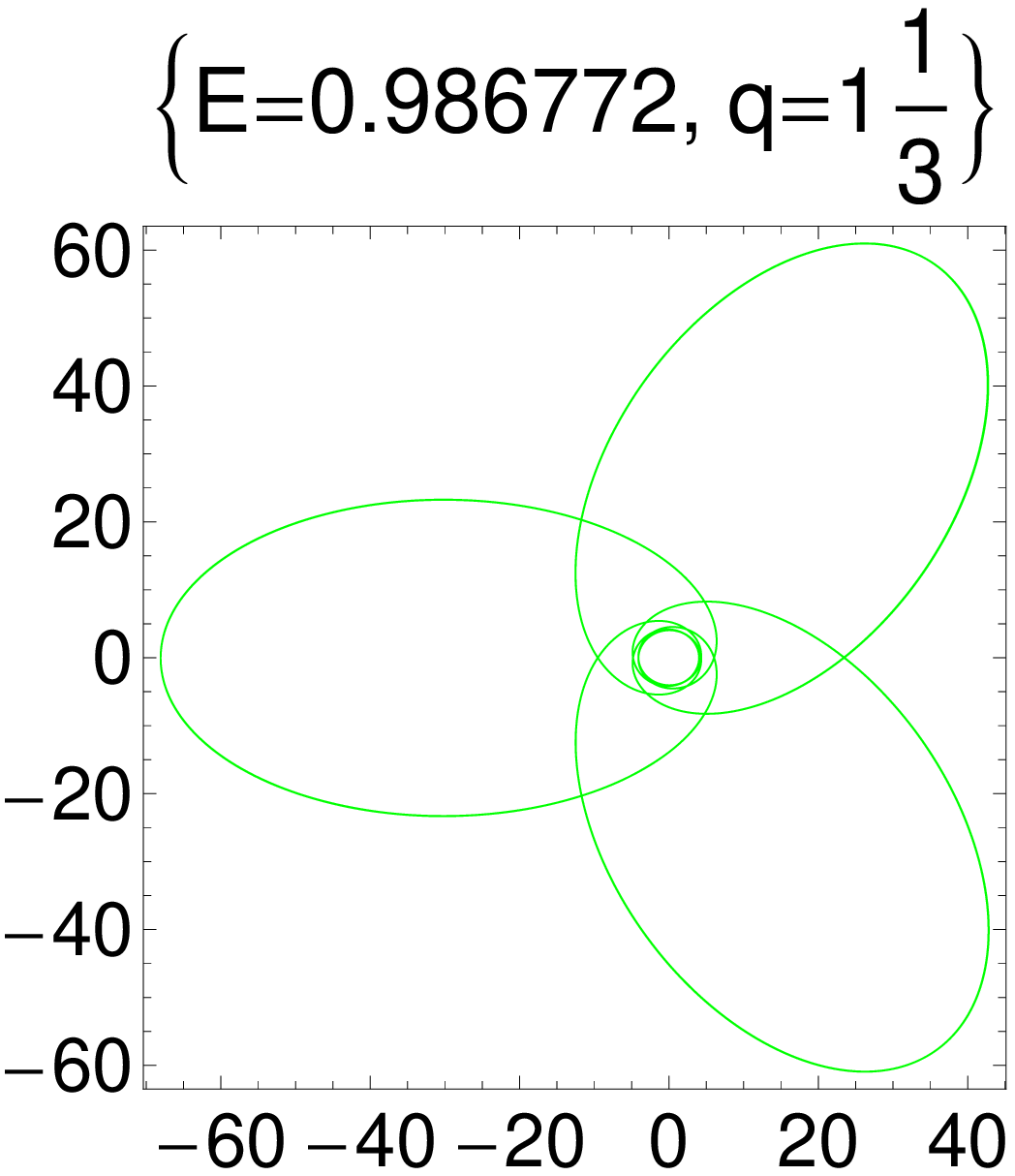}\;\includegraphics[width=4cm]{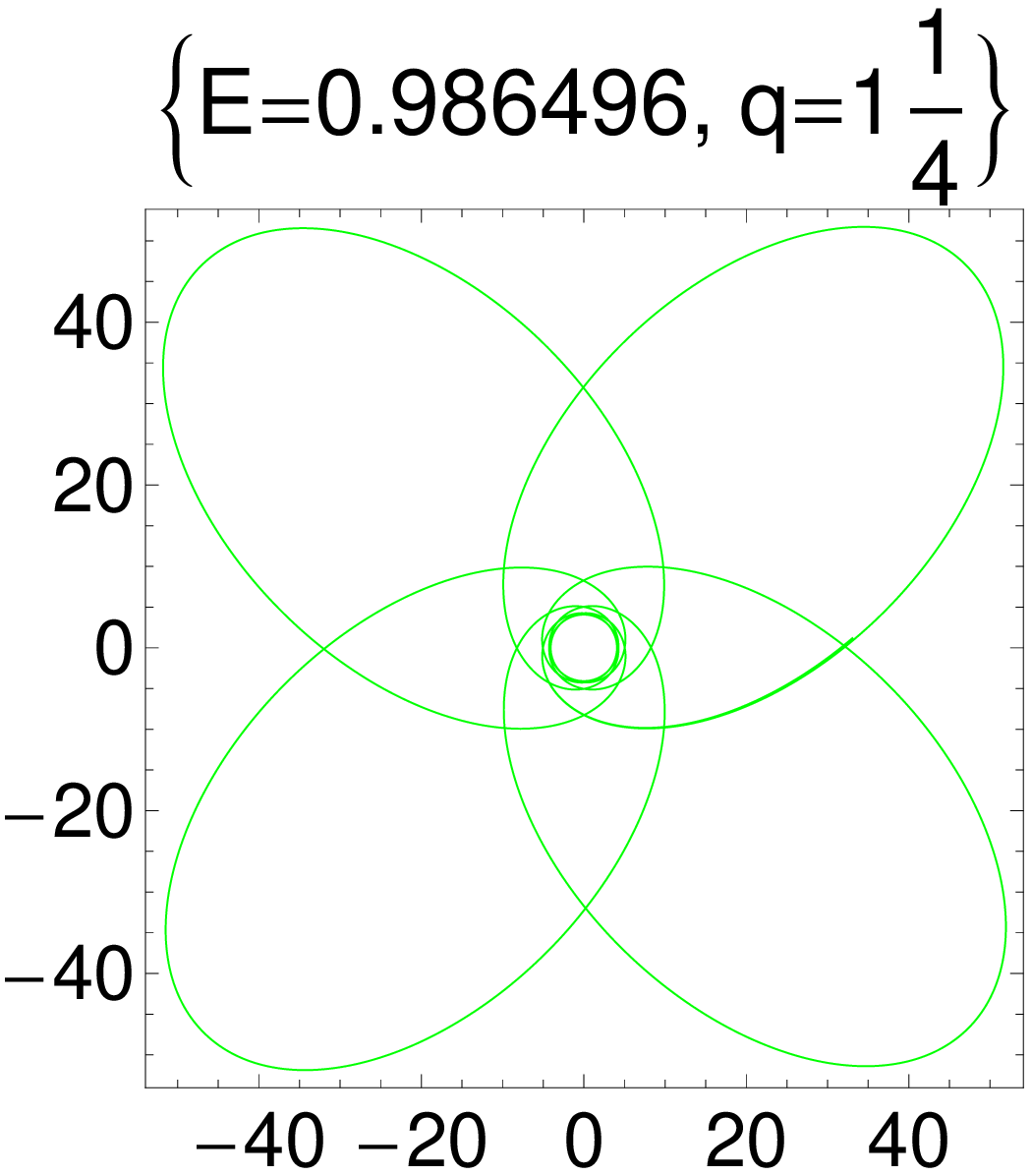}\;
\caption{ Zoom-whirl periodic orbits with $q=w+\frac{v}{z}=1+\frac{1}{z}$, (z = 1, 2, 3, 4) for $a=0$, $b=0.2$, $L=3.8$, here we set $M=1$. \label{figure2}}
\end{center}
\end{figure}
 \begin{figure}[ht]
\begin{center}\includegraphics[width=4cm]{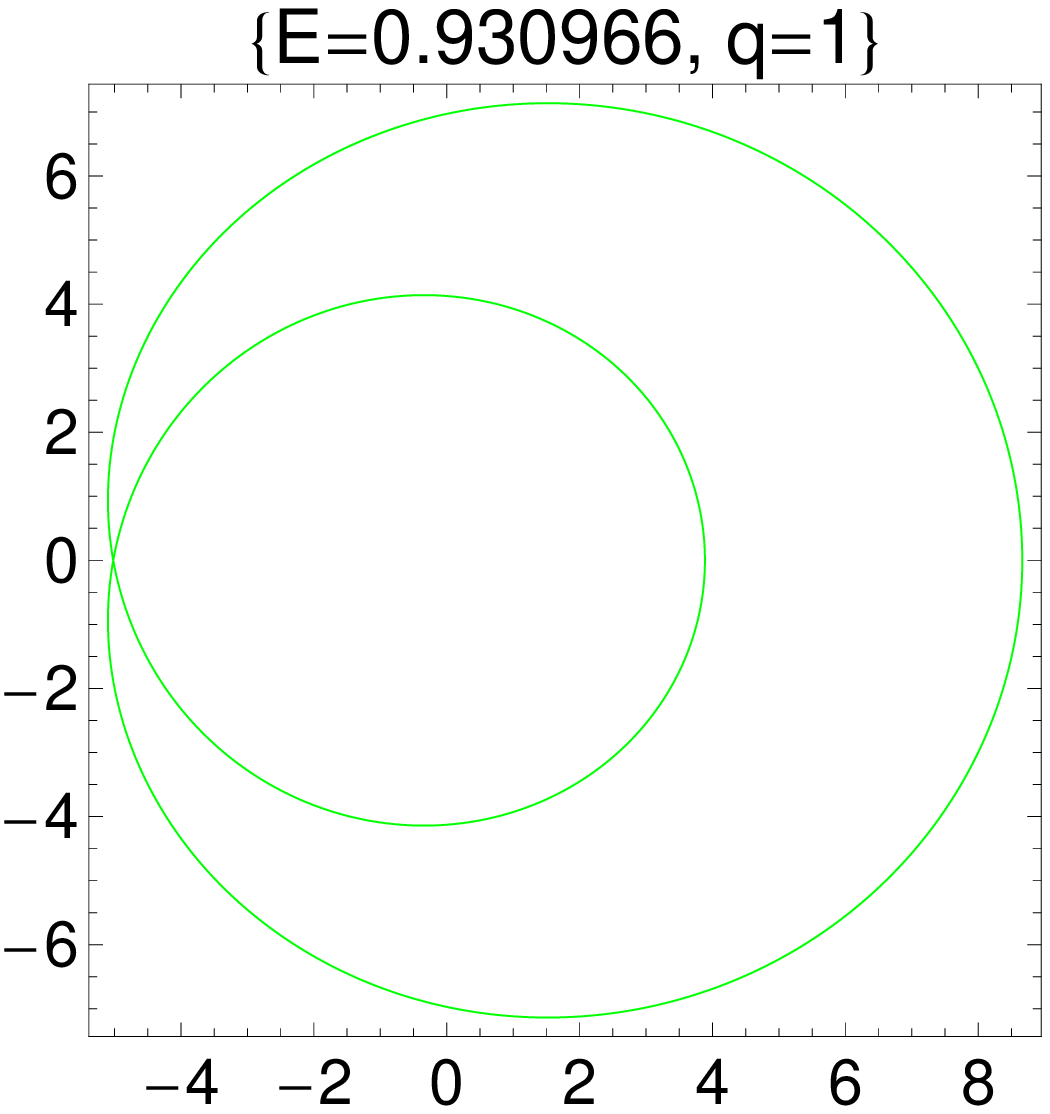}\;\includegraphics[width=4cm]{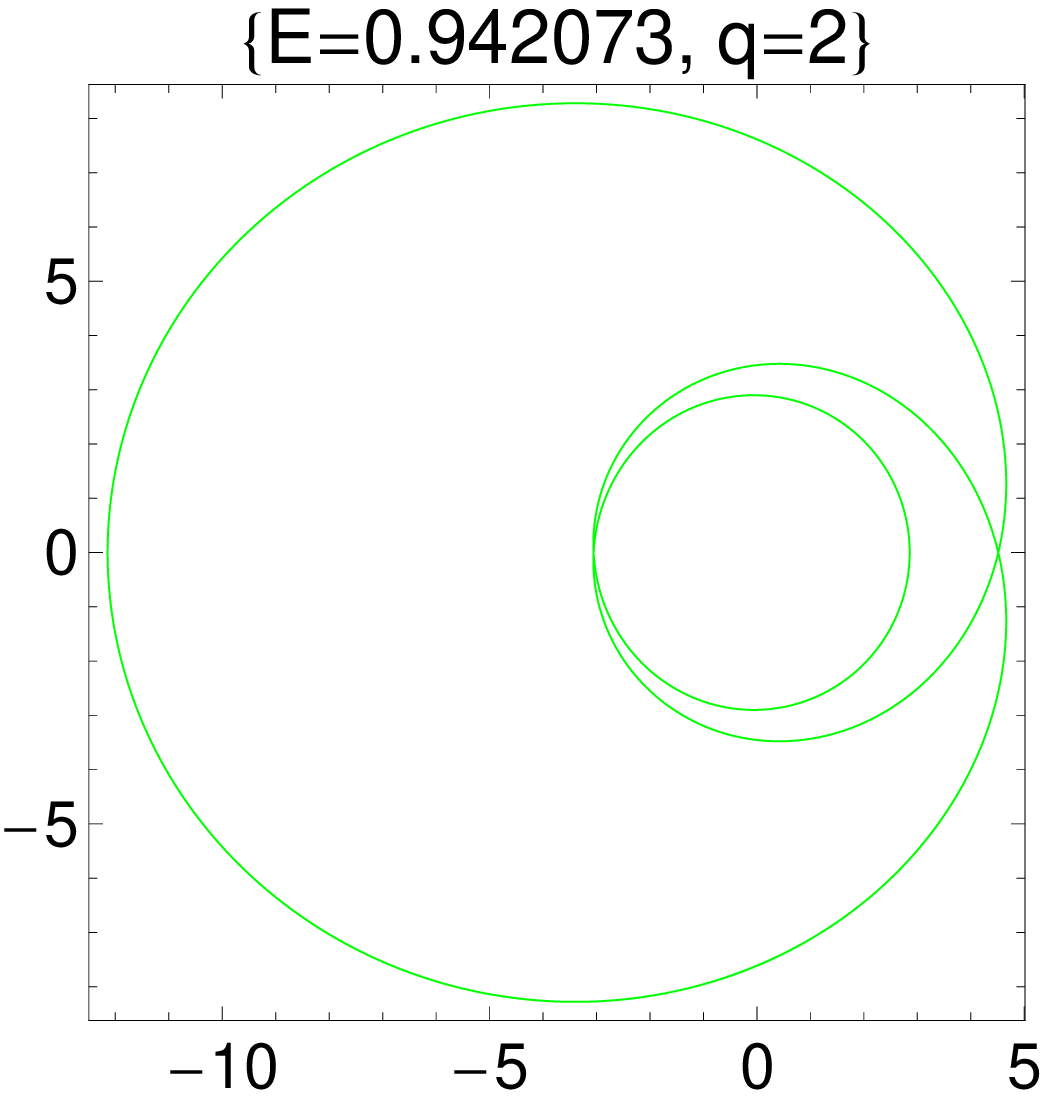}\;\includegraphics[width=4cm]{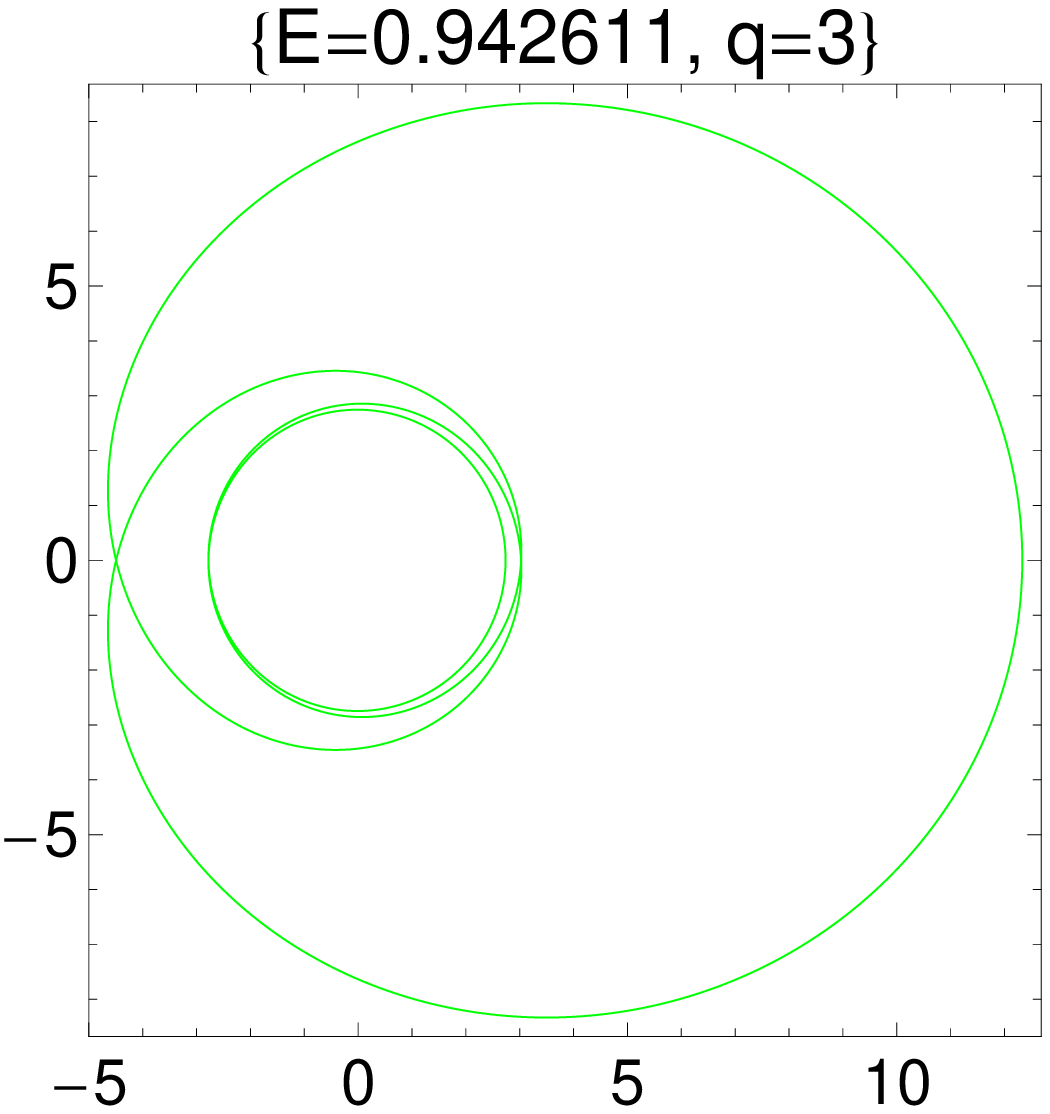}\;
\caption{ Zoom-whirl periodic orbits with $q=w+\frac{v}{z}=w+\frac{0}{1}$, $(w = 1, 2, 3)$ for $a=0.6$, $b=0.1$, $L=2.9$, here we set $M=1$. \label{figure3}}
\end{center}
\end{figure}
 \begin{figure}[ht]
\begin{center}
\includegraphics[scale=0.45]{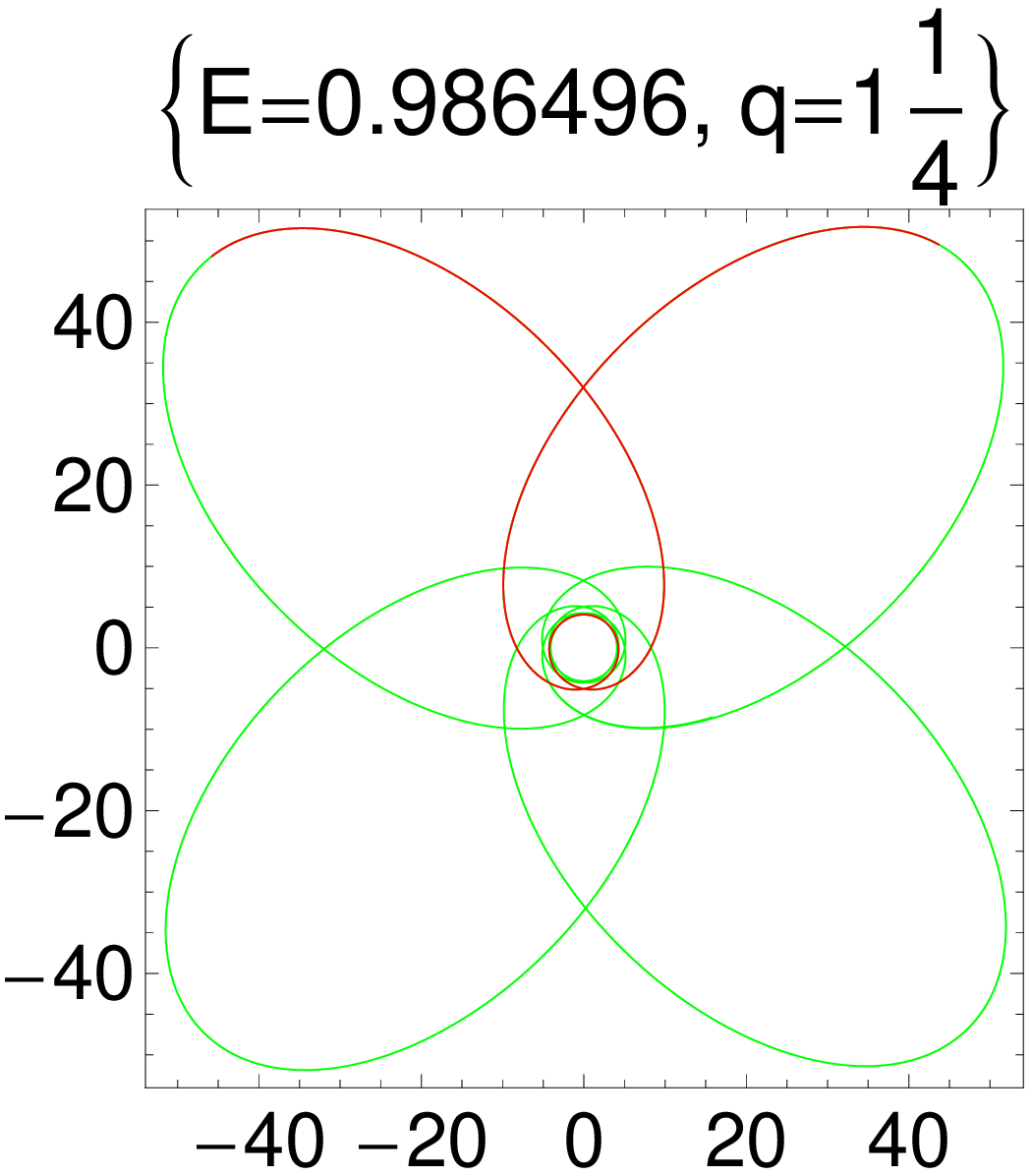},\includegraphics[scale=0.45]{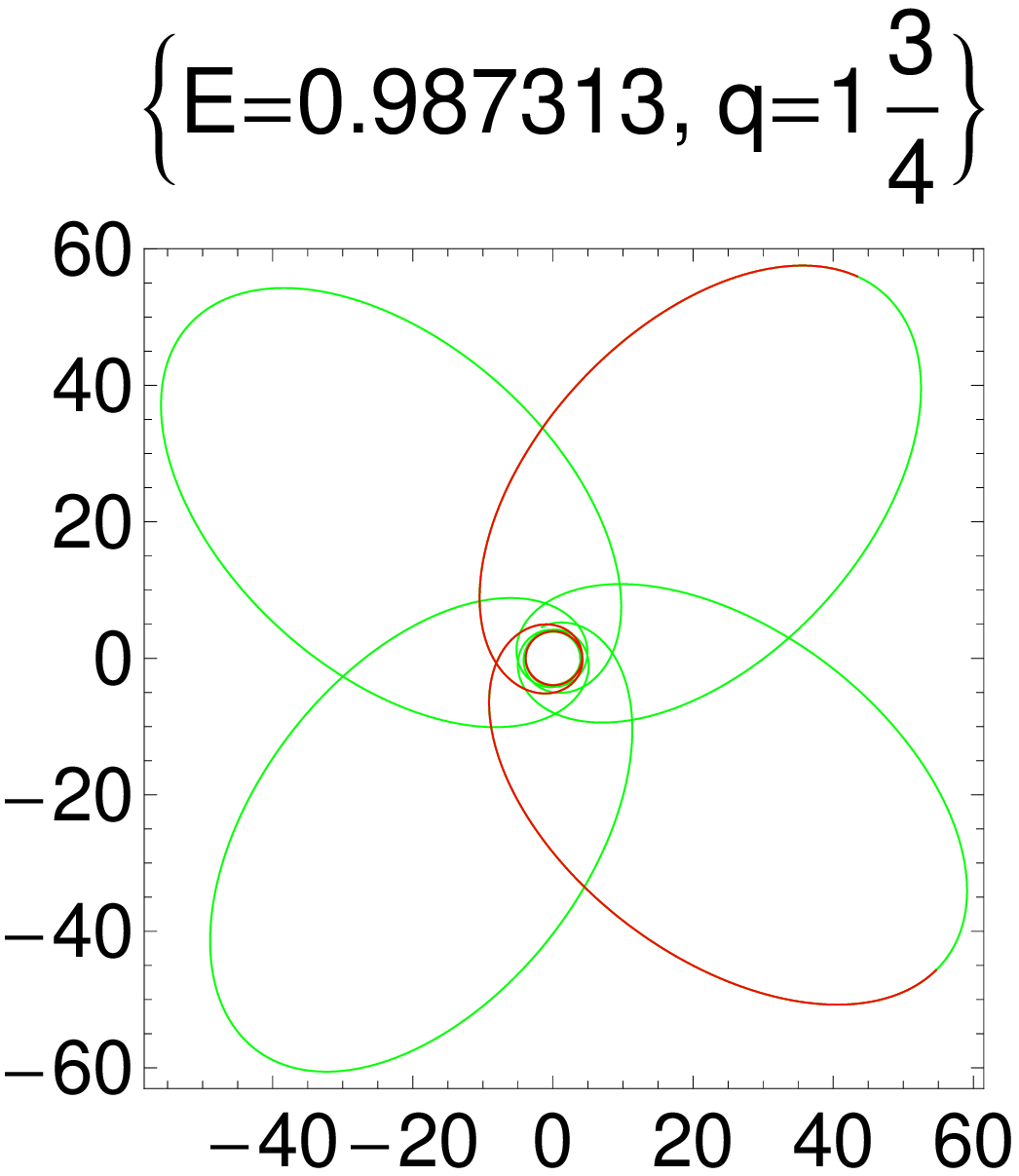},
\caption{ Zoom-whirl periodic orbits with $q=w+\frac{v}{z}=1+\frac{v}{4}$, $(v = 1, 3 )$ for $a=0.$, $b=0.2$, $L=3.8$, here we set $M=1$. \label{figure4}}
\end{center}
\end{figure}

Using the geodesic equations of the KSBH, we get the expression of the rational
number $q$,
\begin{eqnarray}
q&=&\frac{2}{2\pi}\int_{r_p}^{r_a}\frac{\dot \varphi}{\dot r} dr - 1 \quad\\\nonumber&=&\frac{1}{\pi}\int_{r_p}^{r_a}\left (\frac{-\left
(aE-L\right )+\frac{a}{\Delta}((r(r+b)+a^{2})E-aL_{z})}{\sqrt{R}} \right )dr -1 ,\quad
\label{qstart}
\end{eqnarray}
where  $r_p$ and $r_a$ is the periastron and apastron of the zoom-whirl orbit, respectively. In the equatorial plane, one of
the roots is always $0$ and $R$ can be written as
\begin{equation}
R(r)=(E^2-1)r(r-r_0)(r-r_p)(r-r_a).
\end{equation}
Now the rational
number $q$ is a function of $q(a,b,E,L,r_0,r_p,r_a)$. To have $q$ as a function of $(a,b,E,L)$ only, we have to find $r_0,r_p,r_a$ as
functions of $(a,b,E,L)$. Thus we expand the polynomial
$R(r)=(E^2-1)r(r-r_0)(r-r_p)(r-r_a)$ and equate to the definition of $R(r)$
in Eq.\ (\ref{jfc}), matching up coefficients in powers of $r$ and
finding a system of equations for $r_0,r_p,r_a$. Since $r=0$ is always
a root, this is equivalent to a 3rd order equation in $r$
 and cubic
equation have a generic solution. The cubic equation is given as
\begin{equation}Ar^3+Br^2+Cr+D=0\nonumber, \\
\end{equation}
with
\begin{align}
A &= (E^2-1), ~~~~~~~~~~~~~~~~~~~~~~~~~~~
B=2-2b+2E^2b,\nonumber \\
C &=A a^2-L^2+2b-b^2+E^2b^2,~~
D =2(L-aE)^2-a^2bA-L_z^2b.
\label{coeff}
\end{align}
The nonzero roots in ascending order are
\begin{align}
r_0&=-\frac{B}{3A}
-\frac{2^{1/3}(-B^2+3AC)}{3A}(F+\sqrt{G})^{-1/3}
+\frac{1}{3A2^{1/3}}(F+\sqrt{G})^{1/3}
\nonumber, \\
r_p&=-\frac{B}{3A}
+\frac{(1-i\sqrt{3})(-B^2+3AC)}{3A2^{2/3}}(F+\sqrt{G})^{-1/3}
-\frac{1+i\sqrt{3}}{6A2^{1/3}}(F+\sqrt{G})^{1/3}
\nonumber ,\\
r_a&=-\frac{B}{3A}
+\frac{(1+i\sqrt{3})(-B^2+3AC)}{3A2^{2/3}}(F+\sqrt{G})^{-1/3}
-\frac{1-i\sqrt{3}}{6A2^{1/3}}(F+\sqrt{G})^{1/3}
\label{roots},
\end{align}
where
\begin{align}
F&=  -2B^3 + 9ABC-27A^2D\nonumber, \\
G&={F^2-4(B^2-3AC)^3}. \quad\quad
\end{align}
We now have established a simple relationship between rational number $q$ and the quantities $a$, $b$,
$L$ and $E$, by inputting the value of $z,w$ and $v$ for a given $a$, $b$
and $L$ to locate the $E$, apastron $r_a$ and perihelion $r_p$ of the corresponding
periodic orbit.

In Figure \ref{figure2}, we depict zoom-whirl periodic orbits with various $z$ values. When $z$ increases
from $1$ to $4$, the leave of the zoom-whirl periodic orbits varying from one leaf to four leaves.
So ``$z$'' is visualized as the number of leaves, or ``zoom" in the particle orbit.  Figure~\ref{figure3} shows
orbits with various $w$ values, every object travels at least a full from  $4\pi$ around  to $8\pi$ the central
black hole as $w$ increases from  $1$ to $3$. It means that the number of extra turns around the center
of the geometry gives us the value of $w$.  Figure~\ref{figure4} illustrates zoom-whirl orbits with various $v$ values, red line shows
that the zoom-whirl orbits with $v=1$ and $v=3$  move along the different trajectory;  the energy of the zoom-whirl orbit with $v=3$  is higher
than the zoom-whirl orbit with $v=1$.

\begin{figure}[ht]
\begin{center}
\includegraphics[width=4cm]{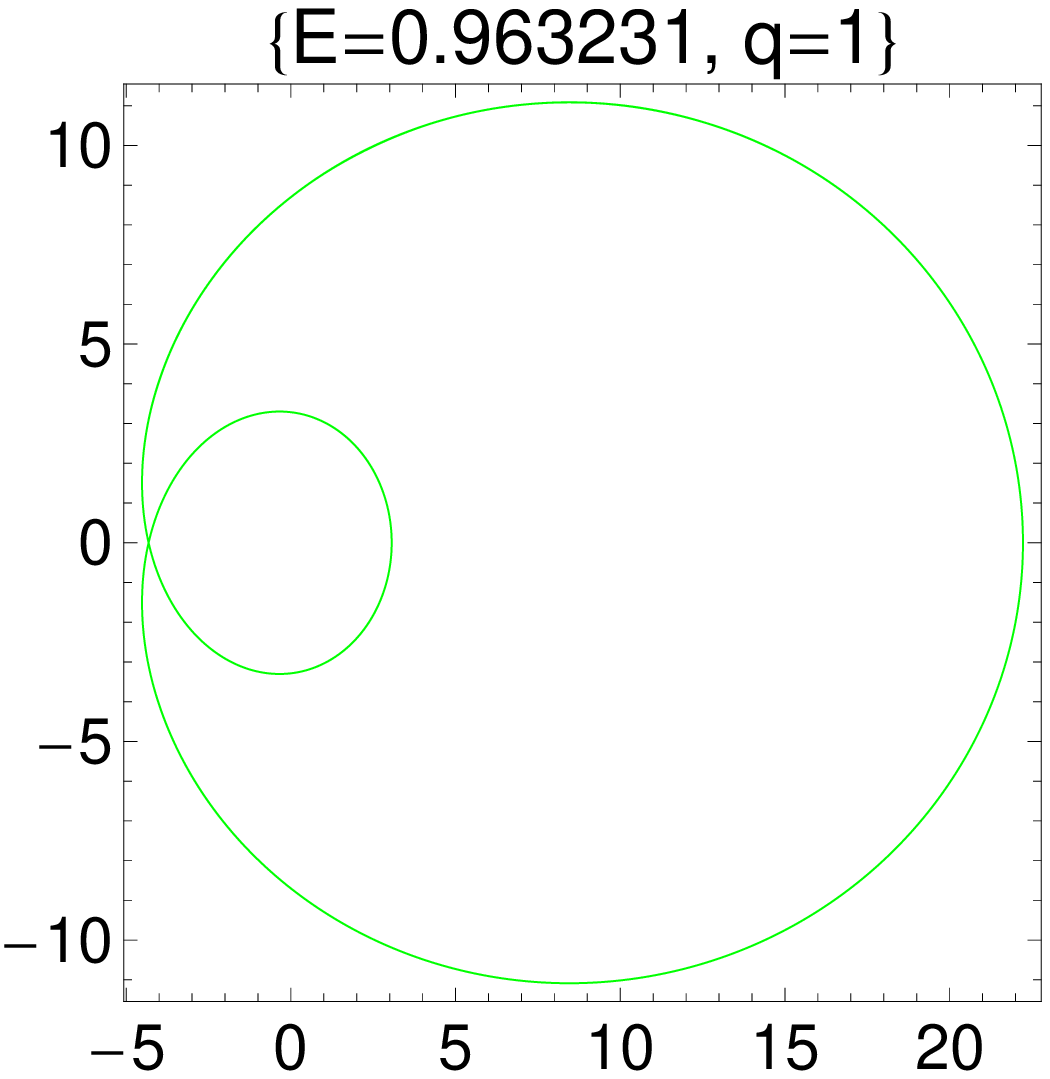}\;\includegraphics[width=4cm]{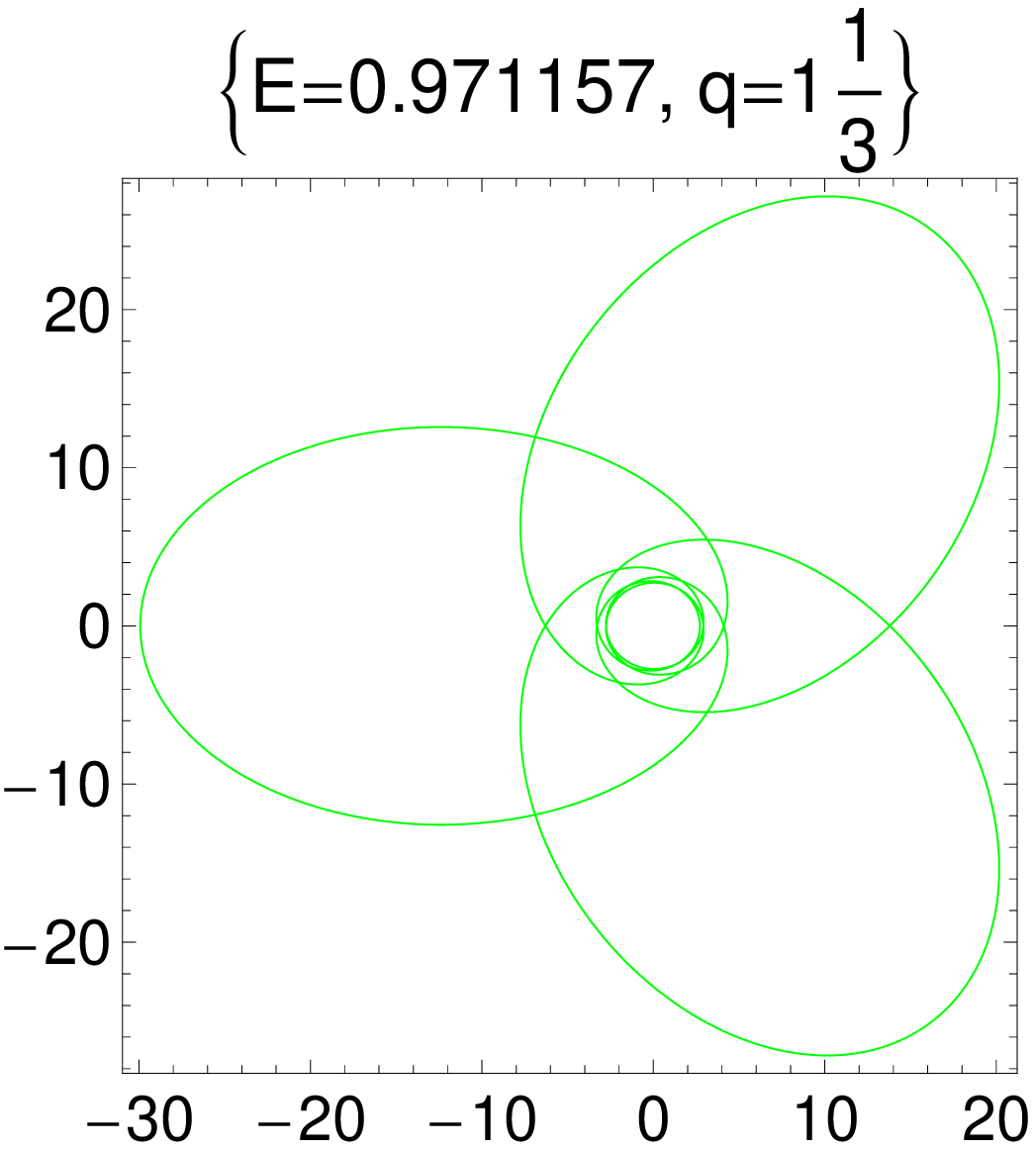}\;\includegraphics[width=4cm]{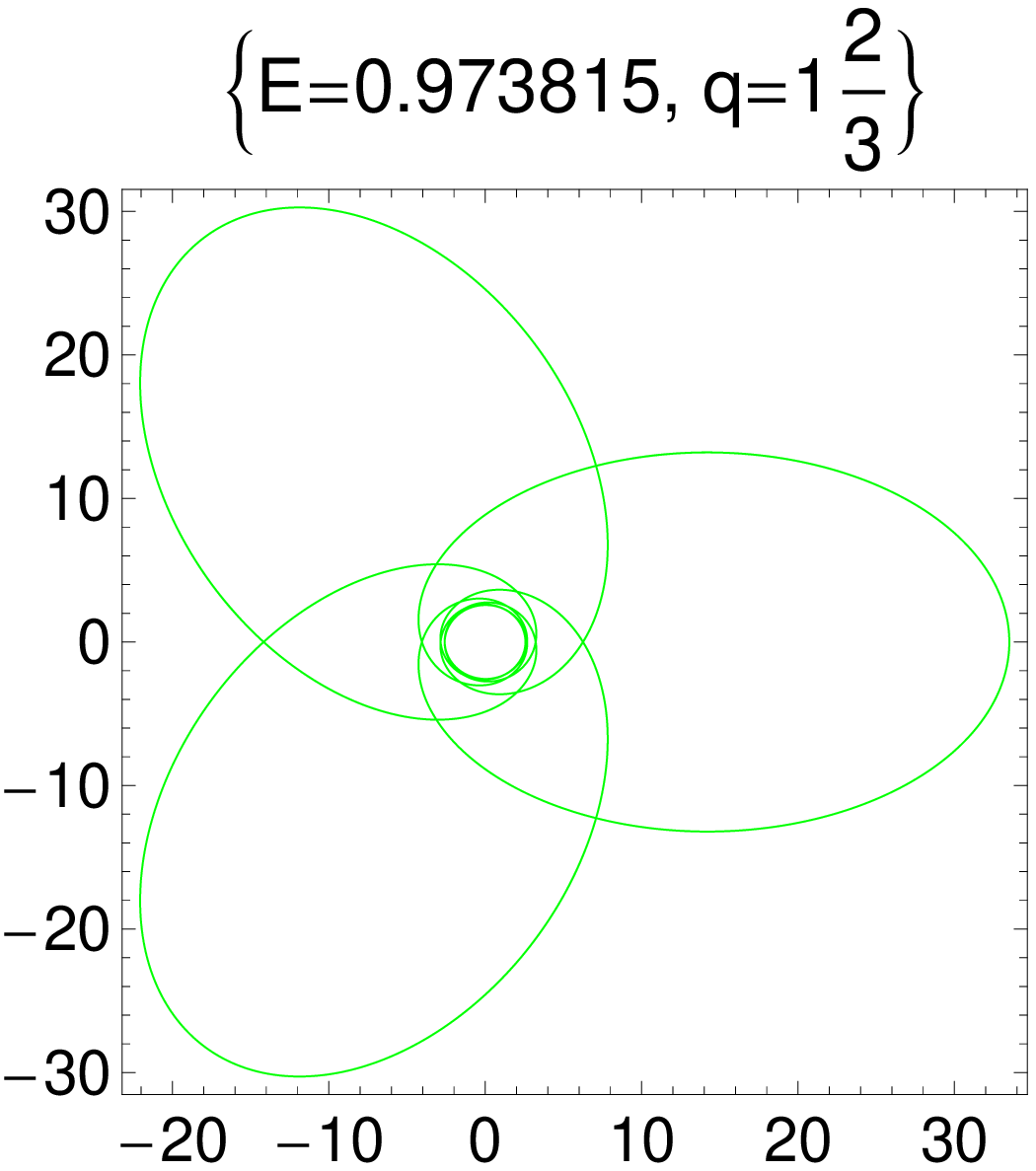}\;
\includegraphics[width=4cm]{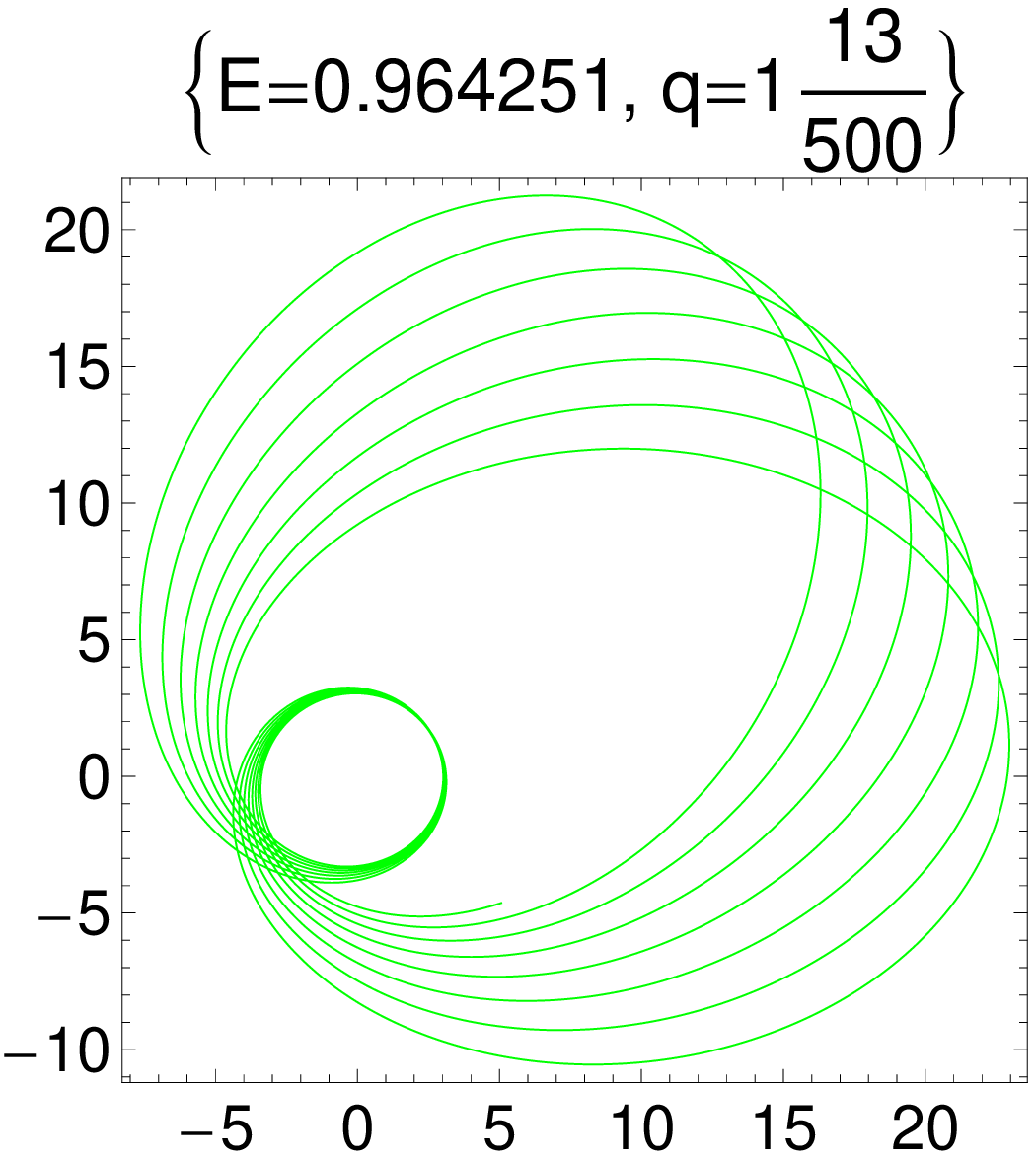}\;\includegraphics[width=4cm]{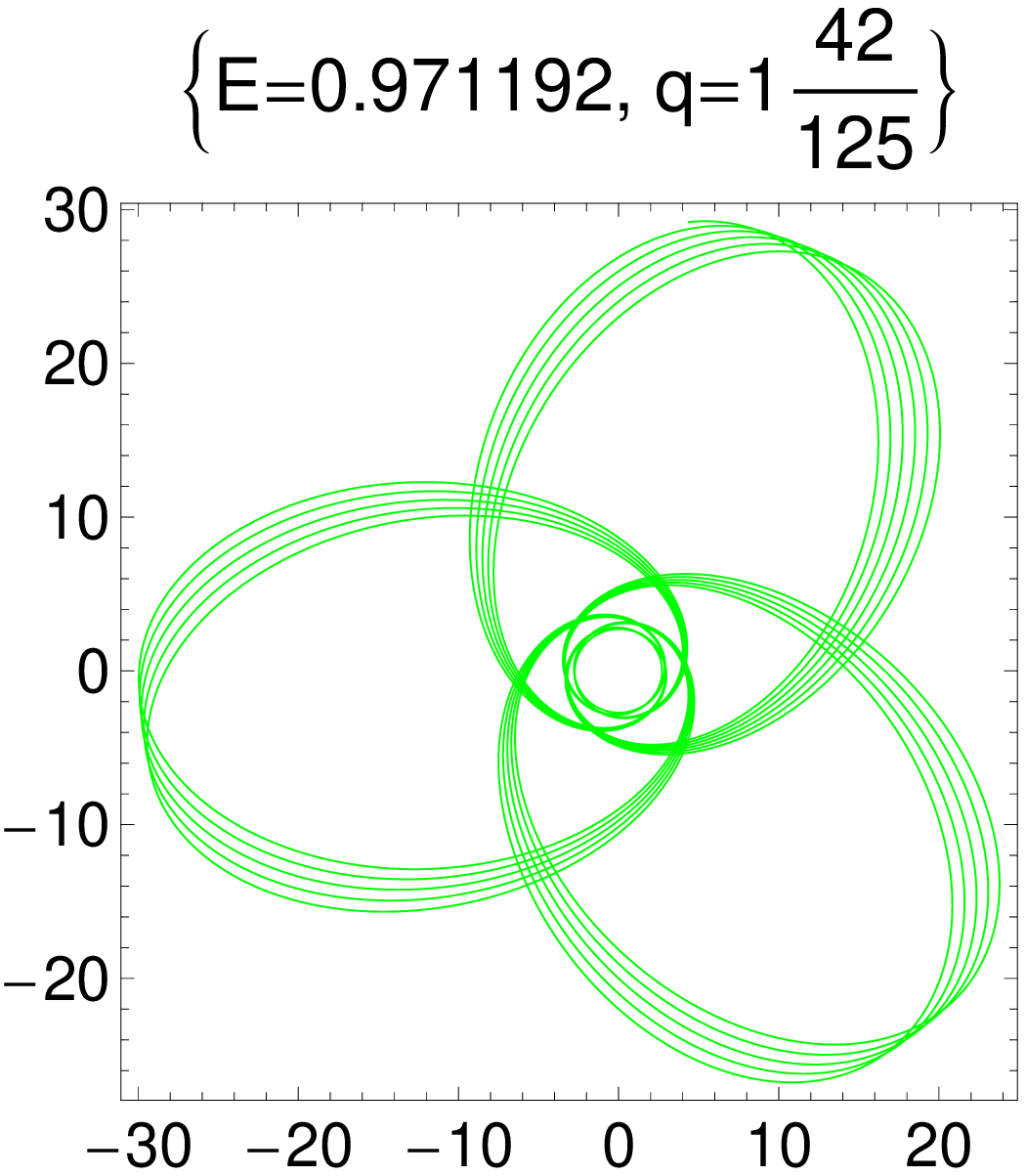}\;\includegraphics[width=4cm]{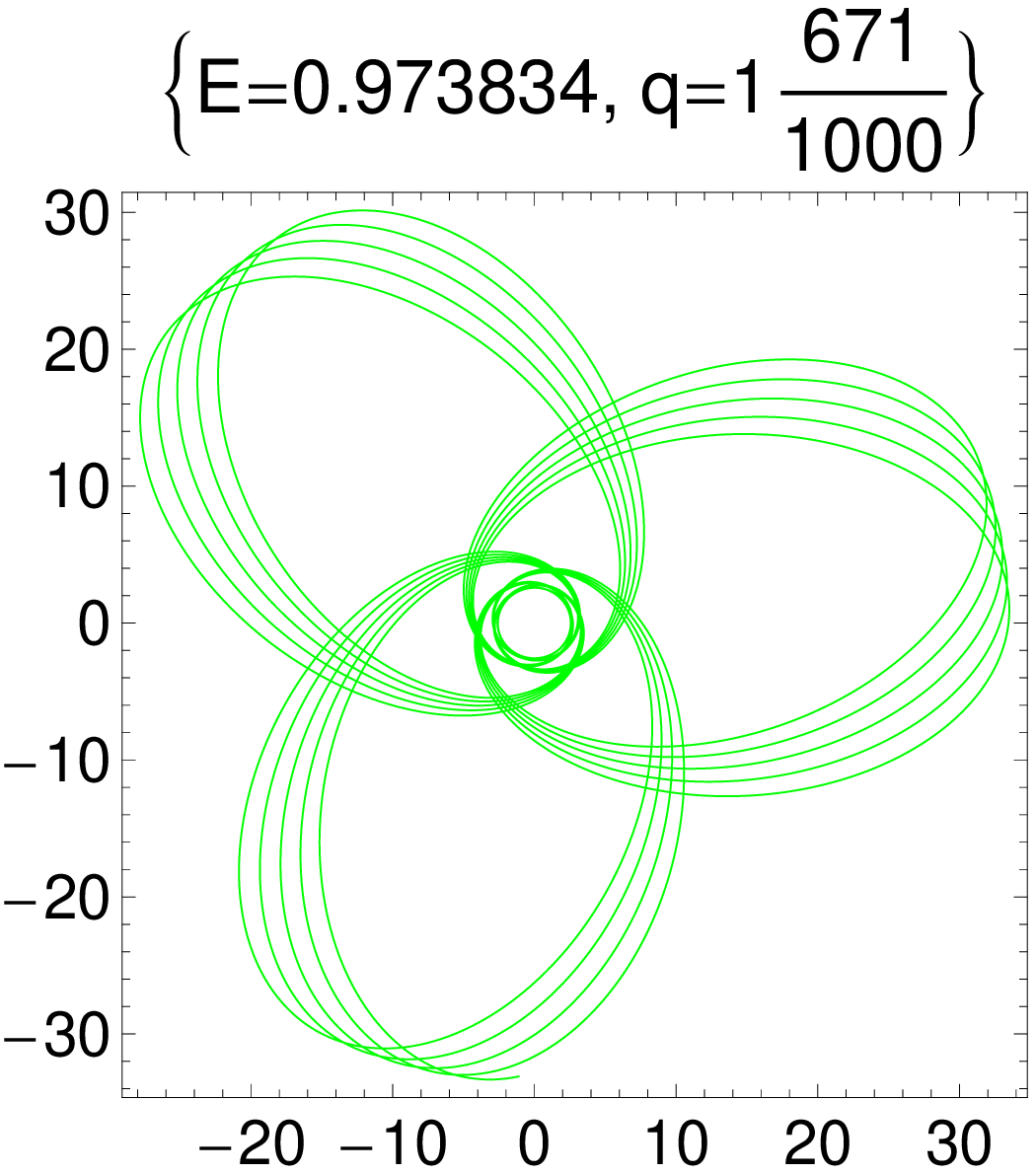}\;
\includegraphics[width=4cm]{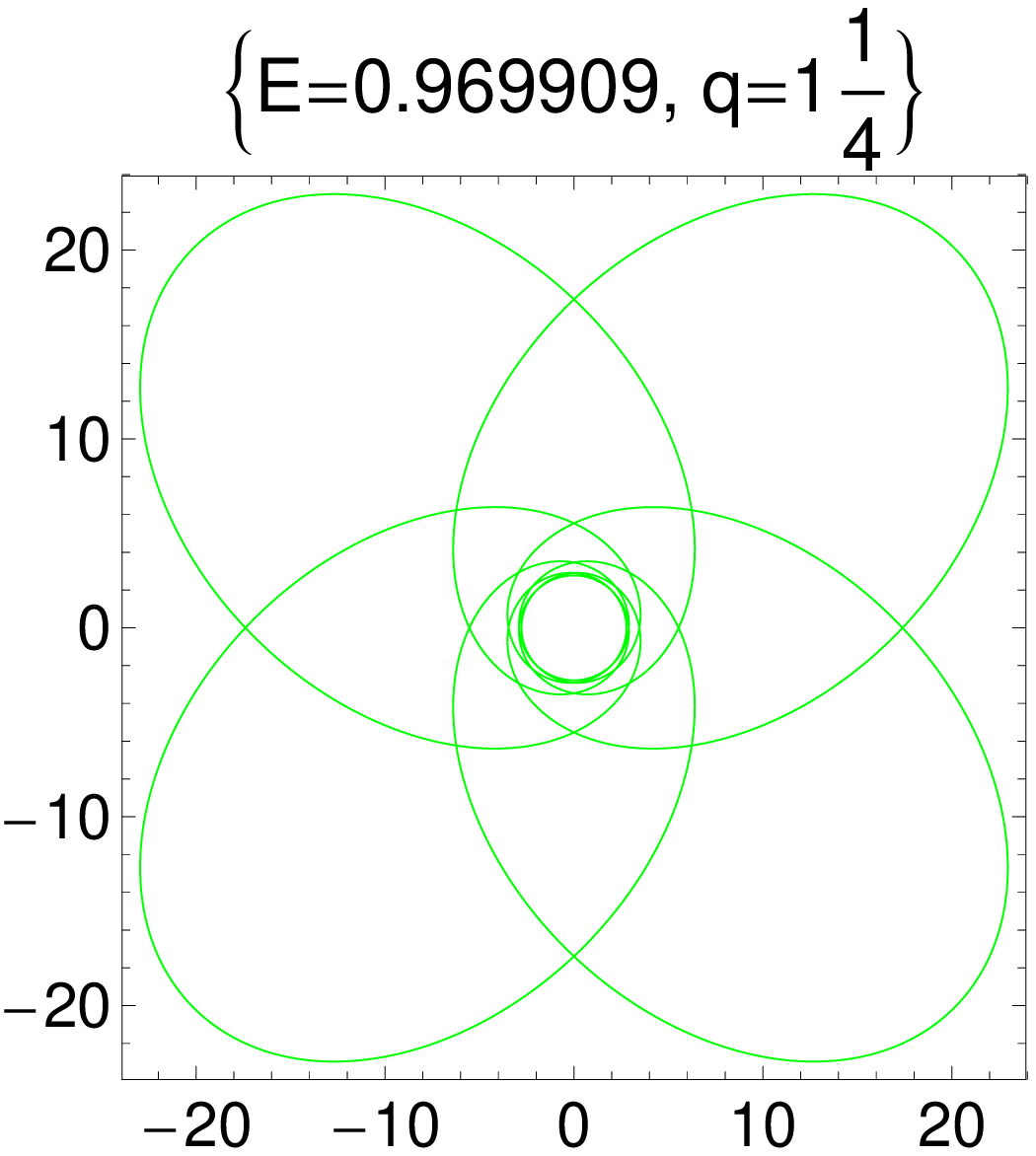}\;\includegraphics[width=4cm]{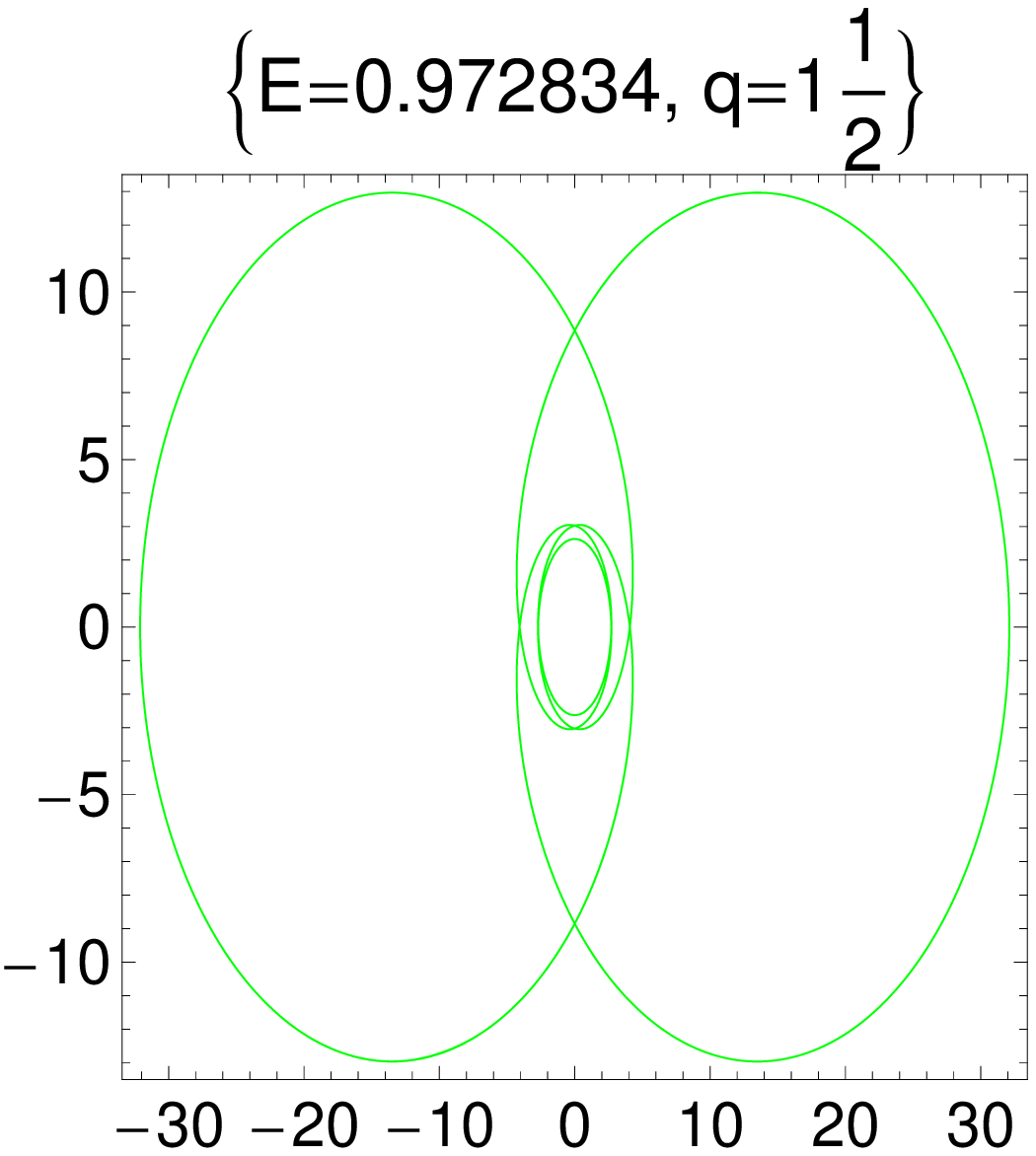}\;\includegraphics[width=4cm]{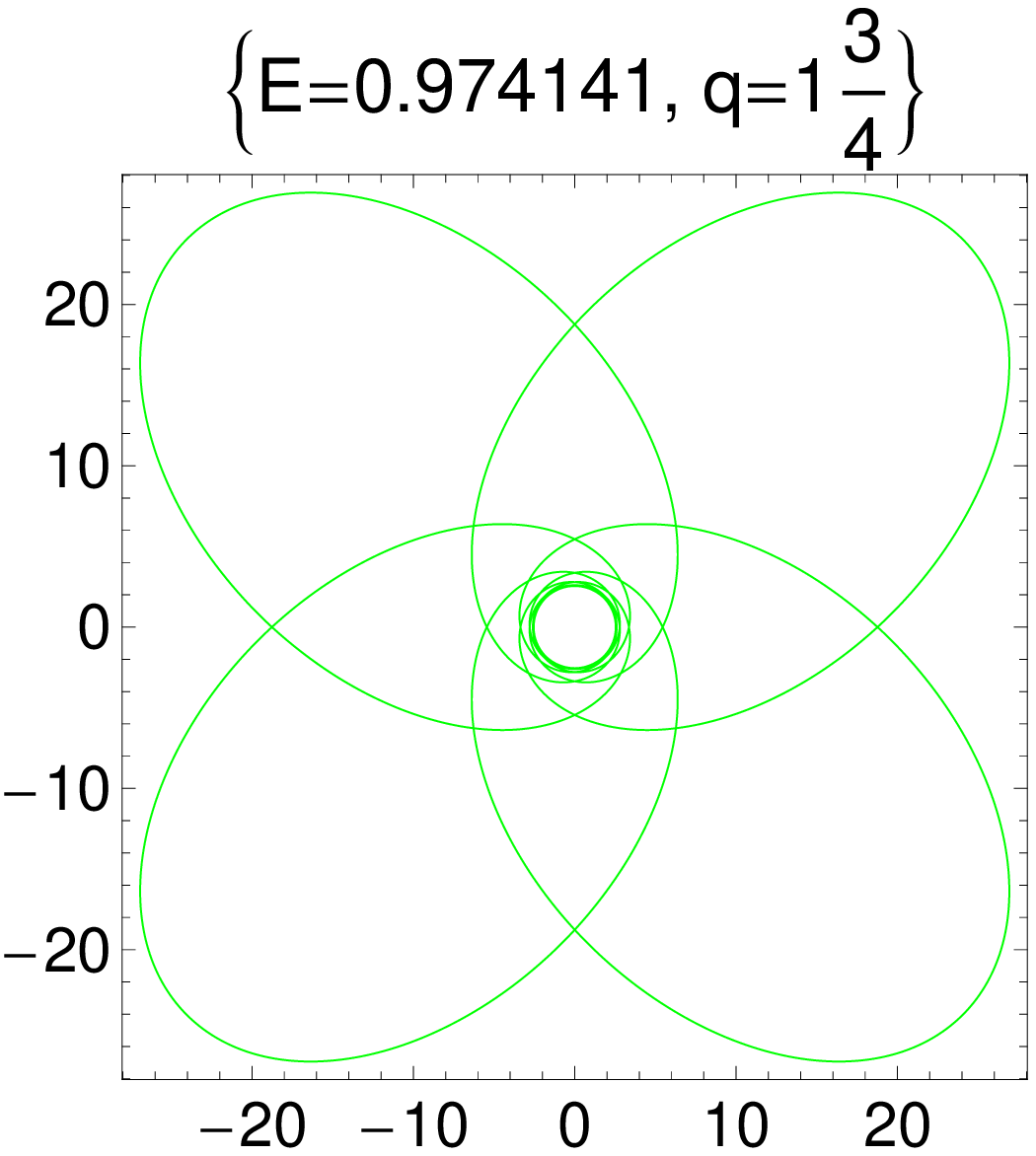}\;
\includegraphics[width=4cm]{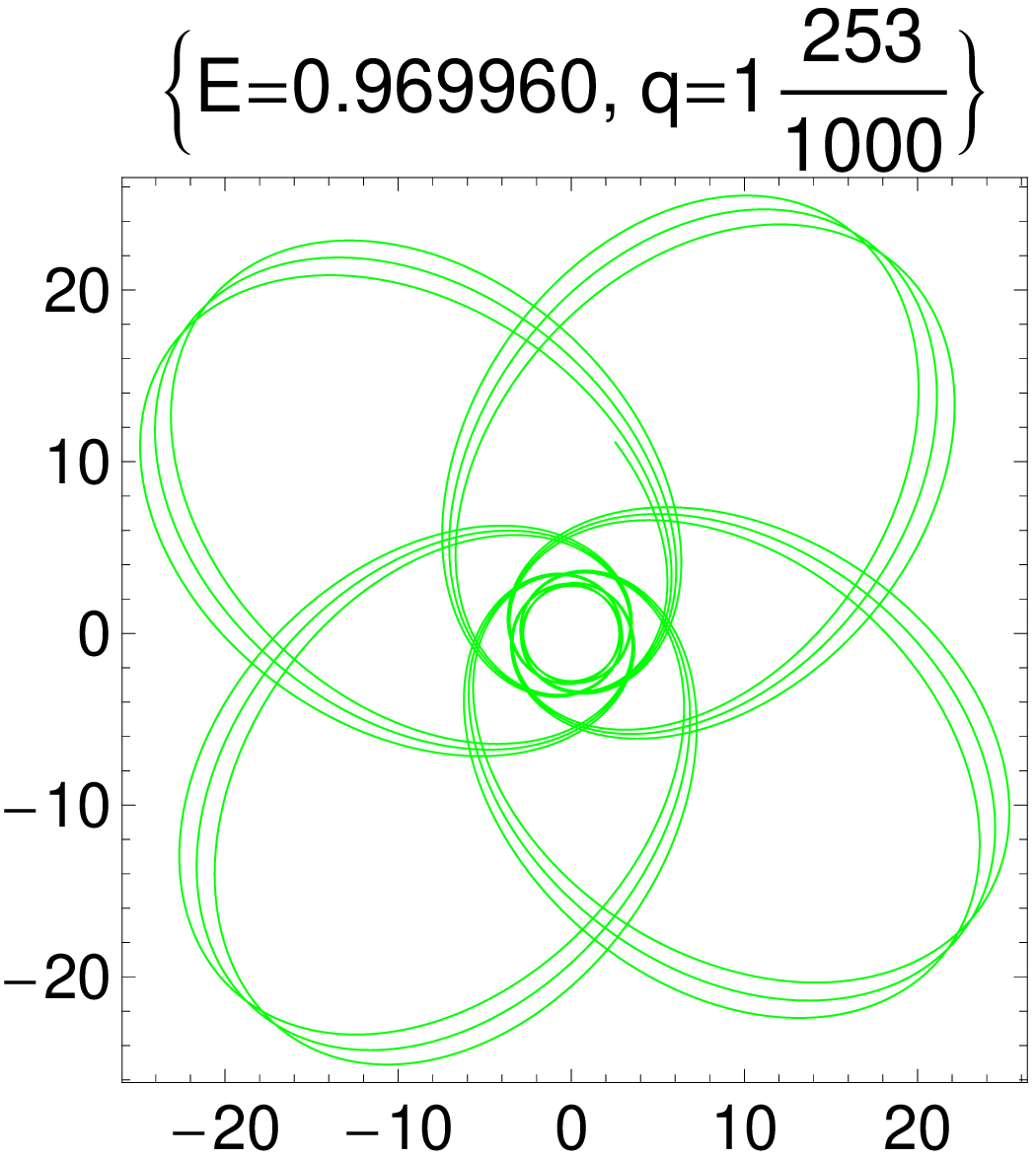}\;\includegraphics[width=4cm]{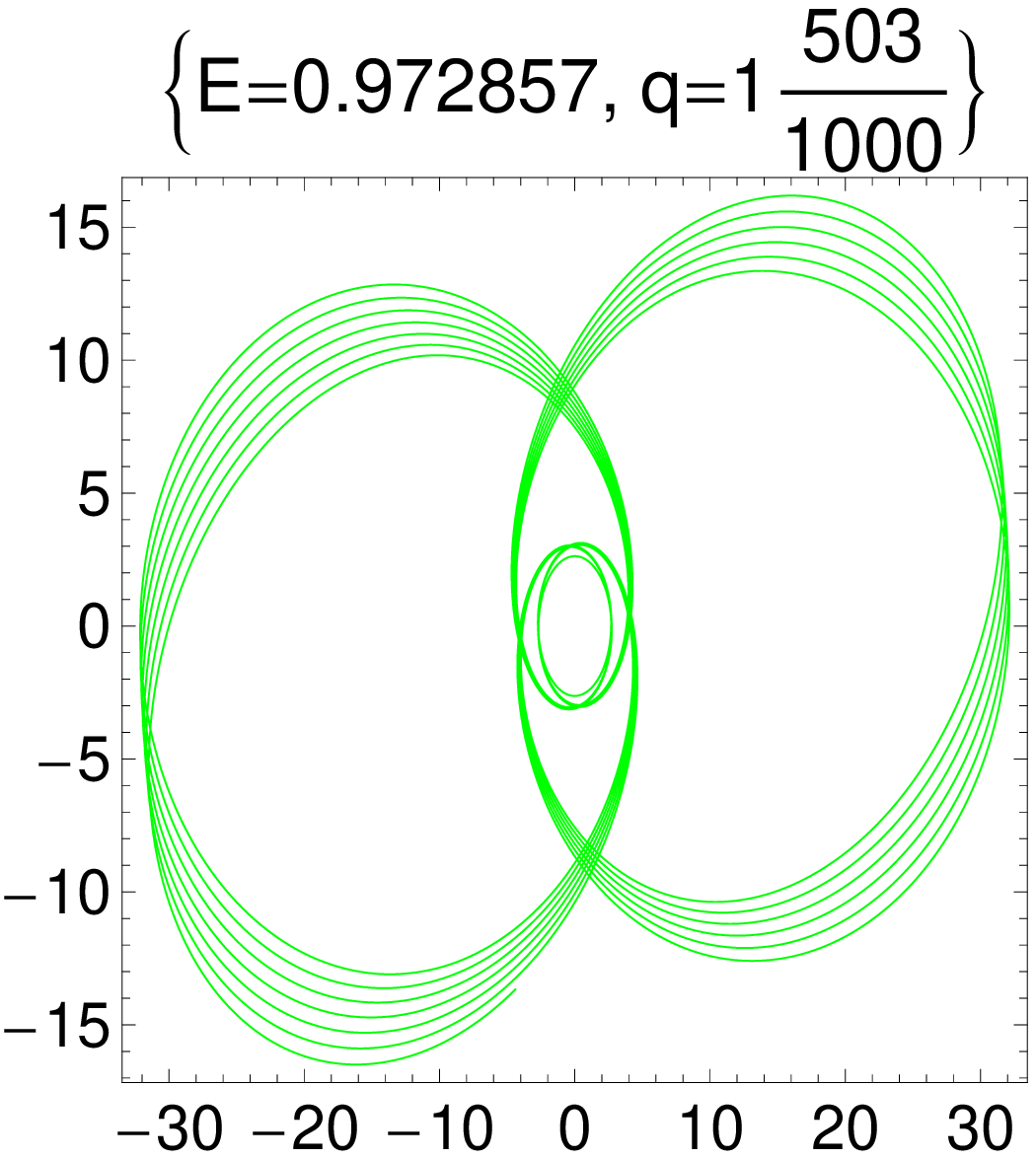}\;\includegraphics[width=4cm]{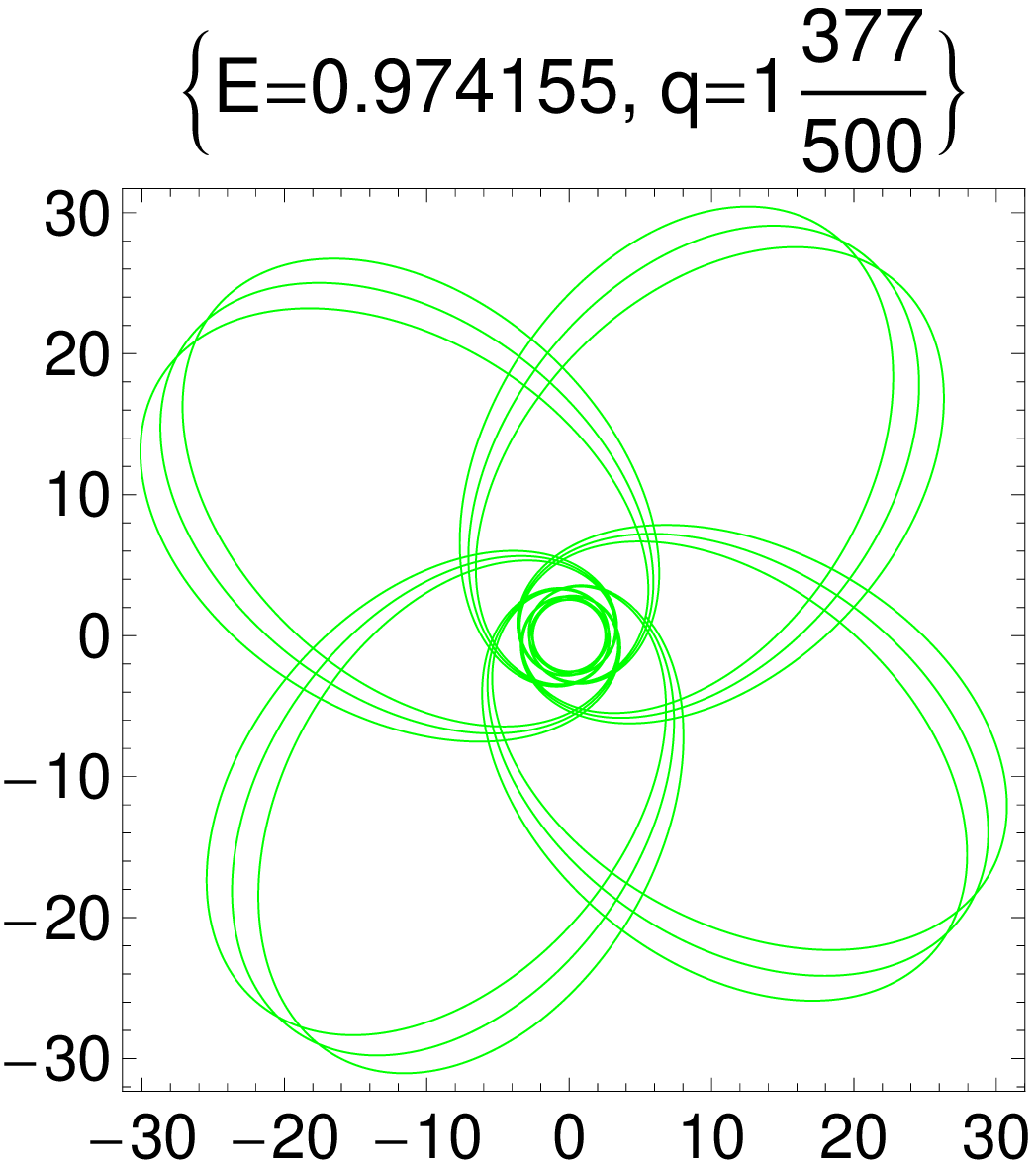}\;
\caption{ A series of  periodic orbits with $a=0.6$, $b=0.18$, $L=3$, $M=1$. Notice that the high $z$ orbits look like precessions of the energetically closest low $z$. \label{figure5}}
\end{center}
\end{figure}

Finally, we must address the degeneracy that arises
when the quotient $v/z$ is a reducible fraction. Thus we require that $v/z$ is an irreducible fraction. As $q$ have approximate values, the zoom-whirl orbit
is precessions. For instance, $v/z =42/125\approx 1/3$, the orbit with $q=167/125$ is the precessions of the orbit with
$q=4/3$. Figure \ref{figure5} shows several pairs of the precession orbits. All orbits with $z = 1,2,3,4$  are drawn. Between each of these low
leaf orbits, randomly selected high zoom orbits are shown as well. The high zoom  orbits ( the second and fourth rows of Fig. \ref{figure5})
look like precessions of the low zoom orbits ( the first and third rows of Fig. \ref{figure5})\cite{Levin}. That is to say, any aperiodic orbit will be arbitrarily well
approximated by a nearby periodic orbit.

\begin{figure}[ht]
\begin{center}
\includegraphics[scale=0.6]{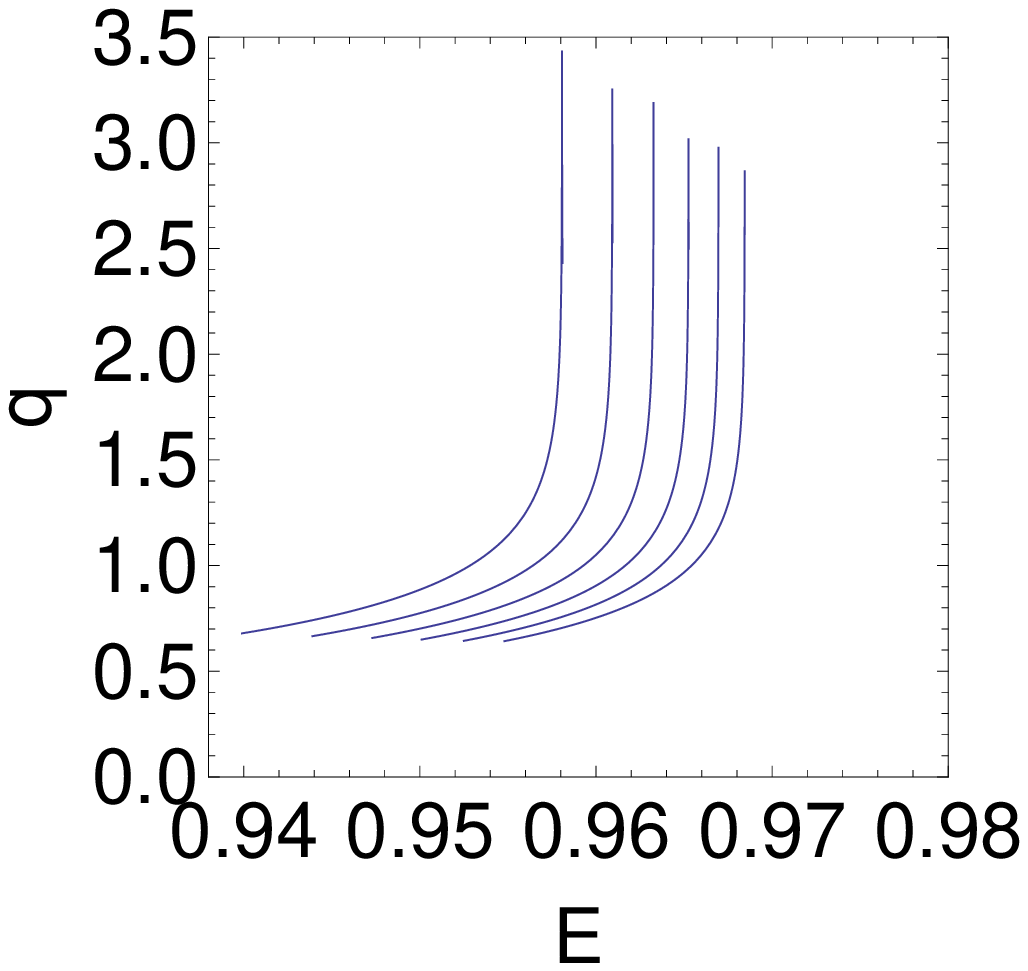},\includegraphics[scale=0.5]{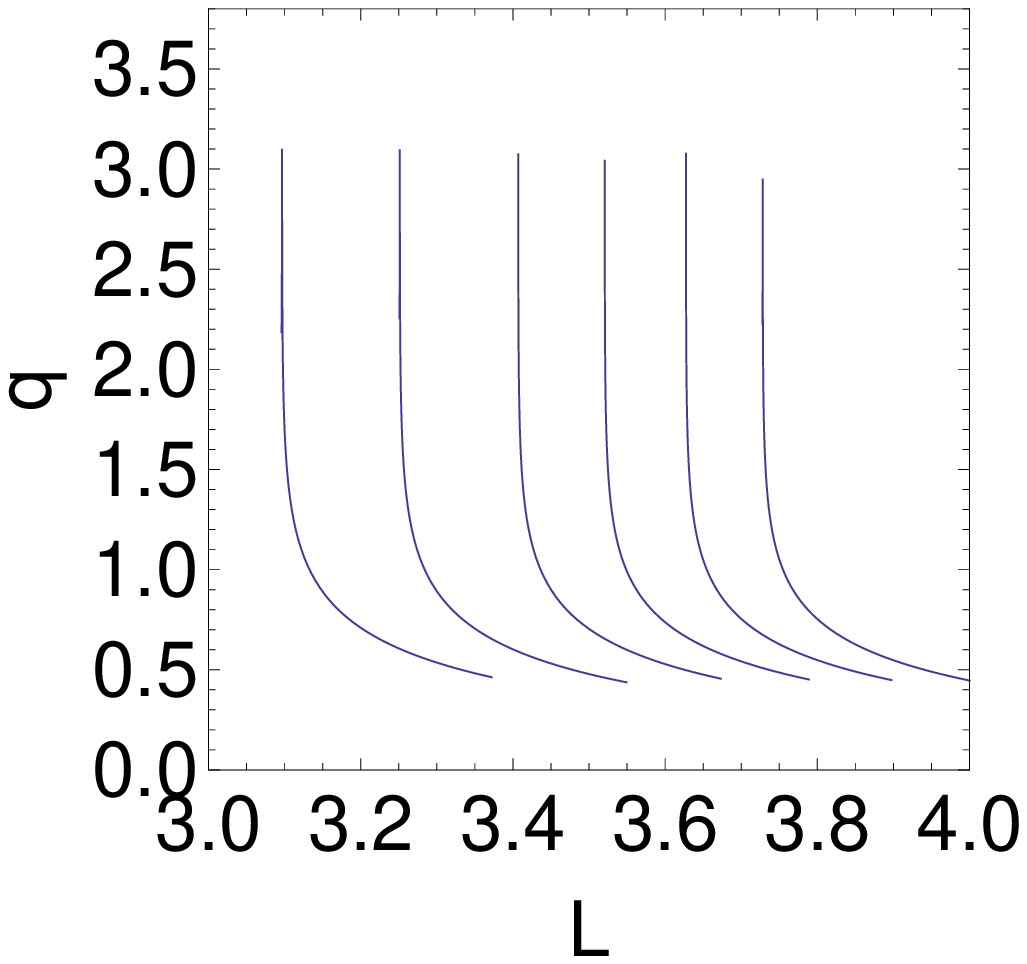},
\caption{The variation of $q$ as a function of energy $E$ and angular momentum $L$ for different values of $b$ in non-rotating KSBH: decreasing from left to right the values are $1, 0.8, 0.6, 0.4, 0.2$, and $0$. In figure (a), for each $b$ depicted above, the corresponding angular momentum is $L_\mathrm{av}=(L_{IBCO}+L_{ISCO})/2$ . In figure (b), energy is kept fixed when $(z,w,v)=(2,1,1)$. \label{figure6}}
\end{center}
\end{figure}
\begin{figure}[ht]
\begin{center}
\includegraphics[scale=0.39]{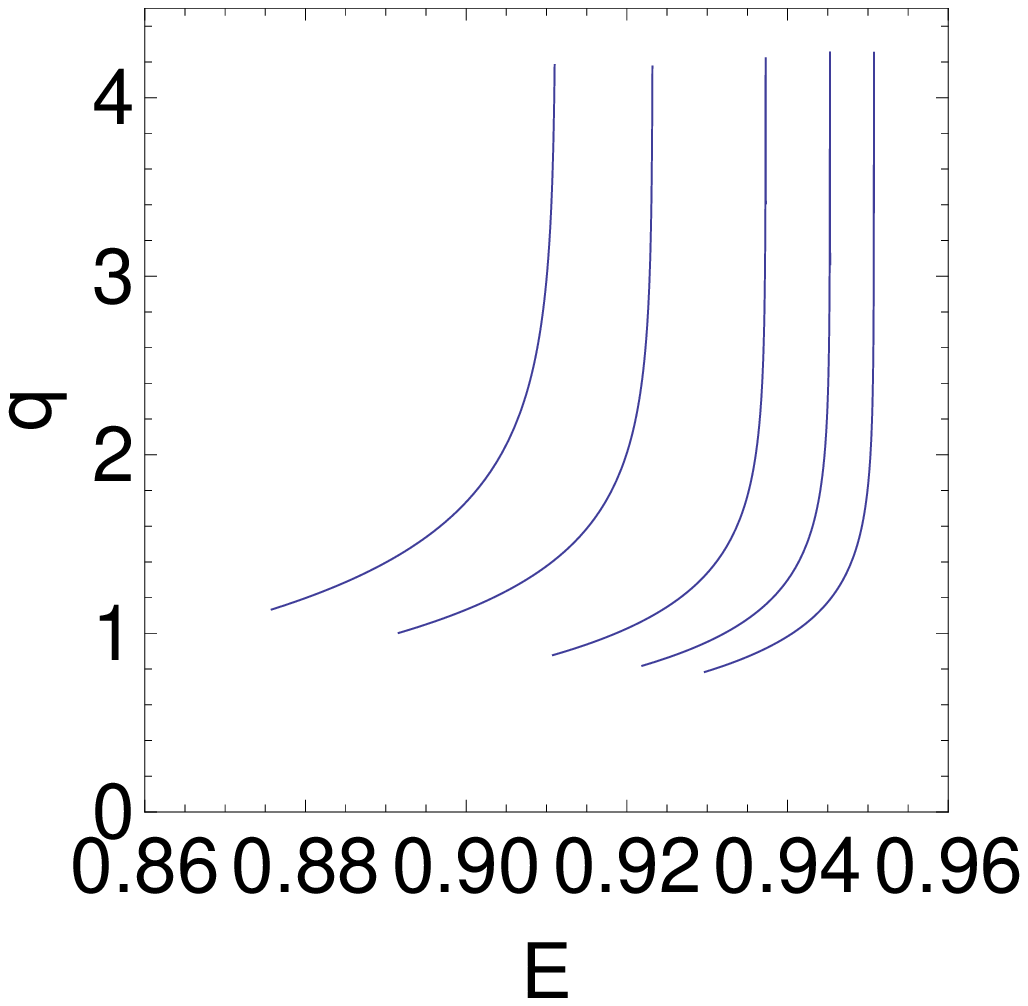},\includegraphics[scale=0.5]{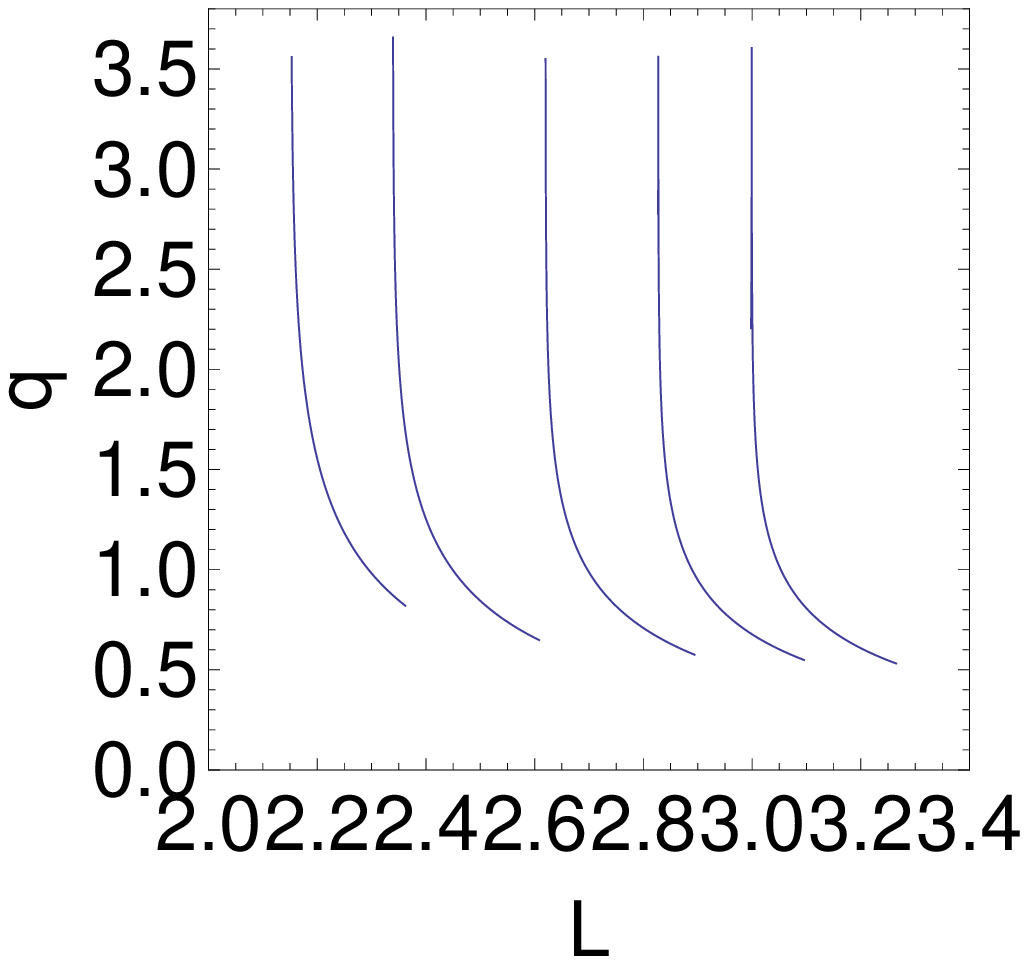},
\caption{The variation of $q$ as a function of energy $E$ and angular momentum $L$ for different values of the charge parameter $b$ in Kerr sen black hole with the fixed rotating parameter $a=0.6$ : decreasing from left to right the values are 0.69, 0.6, 0.4, 0.2, and 0. In figure (a), for each $b$ depicted above, the corresponding angular momentum is  $L_\mathrm{av}=(L_{IBCO}+L_{ISCO})/2$ . In figure (b), energy is kept fixed when $(z,w,v)=(2,1,1)$. \label{figure7}}
\end{center}
\end{figure}
\begin{table}
 \begin{tabular}{ p{1cm} p{2.3cm} p{2.3cm} p{2.3cm} p{2.3cm} p{2.3cm}}
\hline
\textbf{$\nu$} & $L_{\mathrm{av}}$ &$E_{(1,1,0)}$  &$E_{(2,1,1)}$ &$E_{(3,1,1)}$ &$E_{(4,1,1)}$ \\
 \hline
0.    &3.732205   &0.965425          &0.968026              &0.967644          &0.967334 \\
0.2    &3.63135   &0.963682          &0.966493             &0.966076          &0.965739\\
0.4    &3.52488   &0.961665          &0.964732             &0.964272          &0.963902 \\
0.6    &3.41151   &0.959287         &0.962679             &0.962163          &0.961752\\
0.8    &3.28968   &0.956420          &0.960235            &0.959646        &0.959179\\
1      &3.15715   &0.952848        &0.957235             &0.956546          &0.956004\\
\hline
\end{tabular}
\caption{The energy values of $(z=1,2,3,4, w=1, v=1)$ orbits around the non-rotating(a=0) Kerr sen black hole for various $b$ are presented with their corresponding angular momentum, $L_\mathrm{av}=(L_{IBCO}+L_{ISCO})/2$.}\label{tab1}
\end{table}

\begin{table}
 \begin{tabular}{ p{1cm} p{2.3cm} p{2.3cm} p{2.3cm} p{2.3cm} p{2.3cm}}
\hline
\textbf{$\nu$} & $L_{\mathrm{av}}$ &$E_{(1,1,0)}$  &$E_{(2,1,1)}$ &$E_{(3,1,1)}$ &$E_{(4,1,1)}$ \\
 \hline
0.    &3.01045   &0.940589          &0.948703              &0.947290          &0.946238 \\
0.2    &2.84151  &0.932128          &0.942378             &0.940520          &0.939166\\
0.4    &2.63995   &0.918633         &0.932594            &0.929914          &0.928020 \\
0.6    &2.37410   &                 &0.913234             &0.908632         &0.905548\\
0.69    &2.20509   &                &0.893783            &0.887242       &0.883039\\
\hline
\end{tabular}
\caption{The energy values of $(z=1,2,3,4, w=1, v=1)$ orbits around the rotating(a=0.6) Kerr sen black hole for various $b$ are presented with their corresponding angular momentum,  $L_\mathrm{av}=(L_{IBCO}+L_{ISCO})/2$.}\label{tab2}
\end{table}

\section{Energy of generic orbits}\label{ener}

Zoom-whirl periodic orbits give us a way to visually inspect
orbits in different space-time to understand whether we
can distinguish the KSBH from the Kerr black hole. So now we analyze the impact of the charge parameter $b$ on the
zoom-whirl periodic orbits. Transitions in the periodic orbits can be observed when the energy $E$ and angular momentum $L$
changes, which emanate in the form of gravitational waves. Rational number $q$, as a function of $q(a,b,E,L)$, contains
the information on transitions in the periodic orbits during the inspiral stage.  Figures~\ref{figure6} and~\ref{figure7} indicate that
 rational number $q$  monotonically increases with $E$ and decreases with $L$ when the charge parameter $b$ takes the values $1, 0.8, 0.6, 0.4, 0.2$, and $0$  in both the rotating and not-rotating KSBHs. Taking $(z=1,2,3,4, w=1, v=1)$ and  $L_\mathrm{av}=(L_{IBCO}+L_{ISCO})/2$, we list the
 corresponding energy for each periodic orbit in Tables \ref{tab1} and \ref{tab2}. It is shown that the corresponding energy $E$ for each periodic orbit decreases
 with the charge parameter $b$. It implies that the particles in the KSBH with angular momentum $L_{\mathrm{av}}$ in a sufficiently deep potential well  possess a richer variety of bound periodic orbits and a wider range of energy $E$ than their Kerr black hole
counterparts.

\section{summary}

 In this paper, we
have studied periodic orbits in the equatorial plane around the KSBH  with a rational number
 $q$ in terms of three integers $(z,w,v)$ under the taxonomy of orbit of Levin et al. \cite{Levin,Levin1,Levin2,Levin3,Levin4}. By using the Hamiltonian formulation, the geodesic motion of a time-like particle in the KSBH was analyzed, and the bound on the innermost bound and stable circular orbits were also calculated. We found that the angular momentum $L_{ISCO}$ and  $L_{IBCO}$ decreases with the black hole spin $a$ and the charge parameter $b$. We showed that all eccentric
 periodic orbits around the KSBH show zoom whirl behavior of some kind for the angular momentum of the time-like particles in the region  $L_{ISCO}<L<L_{IBCO}$. The characteristic of the
 zoom-whirl periodic orbits is a spectrum of multi-leaf clovers structure, what's
more, aperiodic orbits will look like precessions of
low-leaf clovers in the
strong-field regime. This feature is qualitatively similar to those in the Kerr space-time. Finally, we analyzed the impact of the charge parameter $b$ on the
zoom-whirl periodic orbits to distinguish the KSBH from the Kerr black.
We found that periodic orbits around the KSBH occur at lower energies than their Kerr black hole
counterparts. These results may provide us a possible observational signature by testing
these periodic orbits around the central source to distinguish the KSBH from the Kerr black hole.

\section{\bf Acknowledgments}

Changqing's work was supported by the National Natural Science
Foundation of China under Grant Nos.11447168. Chikun's work was
supported by the National Natural Science Foundation of China under
Grant Nos11247013; Hunan Provincial Natural Science Foundation of
China under Grant Nos. 12JJ4007 and 2015JJ2085.

\end{document}